\documentclass[twocolumn,showpacs,prl,superscriptaddress]{revtex4-1}%
\usepackage{amsfonts}
\usepackage{amsmath}
\usepackage{amssymb}
\usepackage{graphicx}%
\providecommand{\U}[1]{\protect\rule{.1in}{.1in}}

\begin{document}
\title{Engineering nuclear spin dynamics with optically pumped nitrogen-vacancy center}
\author{Ping Wang}
\affiliation{Hefei National Laboratory for Physics Sciences at Microscale and Department of
Modern Physics, University of Science and Technology of China, Hefei, Anhui
230026, China}
\affiliation{Beijing Computational Science Research Center, Beijing 100084, China}
\author{Jiangfeng Du}
\affiliation{Hefei National Laboratory for Physics Sciences at Microscale and Department of
Modern Physics, University of Science and Technology of China, Hefei, Anhui
230026, China}
\author{Wen Yang}
\affiliation{Beijing Computational Science Research Center, Beijing 100084, China}
\email{wenyang@csrc.ac.cn}

\begin{abstract}
We present a general theory for using an optically pumped diamond
nitrogen-vacancy center as a tunable, non-equilibrium bath to control a
variety of nuclear spin dynamics (such as dephasing, relaxation, squeezing,
polarization, etc.) and the nuclear spin noise. It opens a new avenue towards
engineering the dissipative and collective nuclear spin evolution and solves
an open problem brought up by the $^{13}$C nuclear spin noise suppression
experiment [E. Togan \textit{et al}., Nature 478, 497 (2011)].

\end{abstract}

\maketitle

\textit{Introduction.--}Diamond nitrogen-vacancy (NV) center
\cite{GruberScience1997} is a leading platform for quantum computation and
nanoscale sensing
\cite{DuttScience2007,MazeNature2008,DoldeNatPhys2011,ChildressScience2006,ToganNature2010,NeumannNatPhys2010,NeumannScience2010}%
. The NV spin and a few surrounding nuclear spins form a hybrid quantum
register \cite{NeumannScience2008,YaoNatCommun2012,WaldherrNature2014}. Its
coherence time is ultimately limited by the noise from environmental nuclei.
This motivates widespread interest in using the NV spin to control the qubit
and environmental nuclei through their hyperfine interaction (HFI). In
addition to the remarkable success in manipulating
\cite{WaldherrNature2014,TaminiauNatNano2014} a few qubit nuclei, there is
increasing interest in controlling the nuclear spin dissipation, e.g.,
dephasing and relaxation of individual qubit nuclei
\cite{JiangPRL2008,DreauPRL2013} and dynamic polarization of many
environmental nuclei
\cite{KingPRB2010,FischerPRL2013,FischerPRB2013,WangNatCommun2013}. Intensive
experimental efforts have led to dramatic enhancement of the NMR signal
\cite{FischerPRL2013} for applications in chemistry and biomedicine and the
first demonstration of coherence protection by suppressing the nuclear spin
noise \cite{ToganNature2011}. This is an important step towards engineering
the nuclear spin evolution for coherence protection, nanoscale sensing
\cite{MazeNature2008,BalasubramanianNature2008,JonesScience2009,WaldherrNatNano2012}%
, and long-time storage of quantum information
\cite{TaylorPRL2003,TaylorPRL2003a}.

This prospect, however, could be hindered by our limited understanding of the
dissipative nuclear spin dynamics. At present, theoretical treatments are
limited to phenomenological or semiclassical modelling
\cite{JiangPRL2008,JacquesPRL2009,ToganNature2011,DreauPRL2013,WangNatCommun2013}
or numerical simulation neglecting the NV coherence
\cite{FischerPRB2013,FischerPRL2013}. The former provides an intuitive
picture, but is qualitative. The latter is more accurate, but is limited to a
small number of nuclei and may miss important effects due to the NV coherence.
Crucially, it is not clear how to efficiently and quantitatively control the
nuclear spin dissipation and especially the nuclear spin noise, e.g., the
physical mechanism leading to the most impressive observation of Ref.
\cite{ToganNature2011}, the \emph{unconditional} suppression of the $^{13}$C
nuclear spin noise without appreciable polarization, remains unclear.
Subsequent noise suppression experiments \cite{LondonPRL2013,LiuNanoscale2014}
are based on the simple but challenging approach of completely polarizing all
the nuclei or conditioned on measurement-based postselection
\cite{DreauPRL2014}. Despite the recent experimental progress in controlling
the nuclear spin polarization in certain setups
\cite{PaglieroAPL2014,AlvarezArxiv2014}, a general guidance for the efficient,
unconditional control of the nuclear spin dynamics and noise is still lacking.

In this letter, we present a quantum theory for using the NV center to
engineer various nuclear spin dynamics and noise. The essential idea is to
introduce tunable dissipation into the NV center by optical pumping, so the NV
becomes a \emph{tunable}, dissipative bath for the nuclei. When the NV
dissipation is much faster than the NV-induced nuclei dissipation (i.e., the
bath being Markovian), we derive a generalized Lindblad master equation for
the many-nuclei density matrix, with \emph{analytical} expressions for the
nuclei transition/dephasing rates. They not only allow easy calculation of
various nuclear spin dynamics incorporating the NV coherence, but also allow
engineering these dynamics (dephasing, relaxation, squeezing, polarization,
etc.) and the nuclear spin noise by controlling the NV. This is illustrated by
(i) control of nuclear spin relaxation and dephasing, (ii) nuclear spin
squeezing, and (iii) suppression of the noise from many $^{13}$C nuclei. Case
(i) provides a microscopic basis for the phenomenological spin-fluctuator
model \cite{JiangPRL2008} and experimental observations
\cite{DuttScience2007,MaurerScience2012}, and a simple method to suppress the
nuclear spin dephasing or relaxation. Case (iii) provides a general and
efficient way to suppress or amplify the nuclear spin noise and explains the
observed $^{13}$C nuclear spin noise suppression \cite{ToganNature2011} as a
special case.

\textit{General theory.--}We consider many nuclei $\{\hat{\mathbf{I}}_{k}\}$
(described by the Hamiltonian $\hat{H}_{N}$)\ coupled to an optically pumped
NV center. We \emph{always} work in a suitable NV rotating frame and nuclei
interaction picture, so the Hamiltonian consists of the time-independent NV
part $\hat{H}_{e}$, the longitudinal HFI $\hat{K}$ that commutes with $\hat
{H}_{N}$, and the transverse HFI $\hat{V}(t)$ that flips the nuclei:
\begin{equation}
\dot{\rho}(t)=-i[\hat{H}_{e}+\hat{K}+\hat{V}(t),\hat{\rho}(t)]+\mathcal{L}%
_{e}\hat{\rho}(t), \label{EQ0}%
\end{equation}
where $\mathcal{L}_{e}\hat{\rho}\equiv\sum_{fi}\gamma_{fi}\mathcal{D}%
[|f\rangle\langle i|]\hat{\rho}$ is the NV dissipation in the Lindblad form
$\mathcal{D}[\hat{L}]\hat{\rho}\equiv\hat{L}\hat{\rho}\hat{L}^{\dagger}%
-\{\hat{L}^{\dagger}\hat{L},\hat{\rho}\}/2$. Here we focus on NV-induced
nuclei dynamics. The direct nuclei interactions and intrinsic nuclei damping
can be easily included at the end of the derivation.

To derive a closed description for the many-nuclei state $\hat{p}%
(t)\equiv\operatorname*{Tr}_{e}\hat{\rho}(t)$, we use the adiabatic
approximation \cite{CiracPRA1992,WisemanPRA1993a,YangPRB2012} to eliminate the
fast electron motion. We define the many-nuclei basis $|\mathbf{m}\rangle$ as
the common eigenstates of $\hat{H}_{N}$ and $\hat{K}$ with $\hat{K}%
|\mathbf{m}\rangle=\hat{K}_{\mathbf{m}}|\mathbf{m}\rangle$, where $\hat
{K}_{\mathbf{m}}$ is an electron operator. The block $\hat{\rho}%
_{\mathbf{m},\mathbf{n}}\equiv\langle\mathbf{m}|\hat{\rho}|\mathbf{n}\rangle$
obeys
\[
\dot{\rho}_{\mathbf{m},\mathbf{n}}=\mathcal{L}_{\mathbf{m},\mathbf{n}}%
\hat{\rho}_{\mathbf{m},\mathbf{n}}-i\{\hat{\rho}_{\mathbf{m},\mathbf{n}%
},\delta\hat{K}_{\mathbf{m,n}}\}/2-i\langle\mathbf{m}|[\hat{V},\hat{\rho
}]|\mathbf{n}\rangle,
\]
where $\delta\hat{K}_{\mathbf{m,n}}\equiv\hat{K}_{\mathbf{m}}-\hat
{K}_{\mathbf{n}}$ and $\mathcal{L}_{\mathbf{m},\mathbf{n}}(\bullet
)\equiv-i[\hat{H}_{e}+(\hat{K}_{\mathbf{m}}+\hat{K}_{\mathbf{n}}%
)/2,\bullet]+\mathcal{L}_{e}(\bullet)$. Tracing over the electron yields
\[
\dot{p}_{\mathbf{m},\mathbf{n}}=-i\operatorname*{Tr}\nolimits_{e}\{\hat{\rho
}_{\mathbf{m},\mathbf{n}},\delta\hat{K}_{\mathbf{m,n}}\}/2-i\operatorname*{Tr}%
\nolimits_{e}\langle\mathbf{m}|[\hat{V},\hat{\rho}]|\mathbf{n}\rangle
\]
for $p_{\mathbf{m},\mathbf{n}}\equiv\langle\mathbf{m}|\hat{p}|\mathbf{n}%
\rangle$. The above equations contain three dissipation time scales:\ NV
dissipation (time scale $T_{e}$) driven by $\mathcal{L}_{\mathbf{m}%
,\mathbf{n}}$, nuclei dephasing (time scale $T_{2}$) by $\delta\hat
{K}_{\mathbf{m},\mathbf{n}}$ fluctuation, and nuclei relaxation (time scale
$T_{1}$) by $\hat{V}(t)$ fluctuation. Nuclei dissipation much slower than
$T_{e}$ can be adiabatically singled out. For specificity, we consider
$T_{e}\ll T_{1},T_{2}$ and single out all the dynamics of $\hat{p}(t)$ on the
coarse grained time scale $\Delta t\gg T_{e}$.

By treating $\mathcal{L}_{\mathbf{m},\mathbf{n}}$ exactly and $\delta\hat
{K}_{\mathbf{m,n}}$, $\hat{V}(t)$ perturbatively, application of the adiabatic
approximation \footnote{See supplementary material for derivation of Eqs.
(2)--(6) (Sec. I), a perturbative, explicit expression for $W_{\mathbf{p}%
\leftarrow\mathbf{m}}$ (Sec. II), and summary of the NV Hamiltonian,
NV-induced nuclear spin transition rates, and calculation of NV fluorescence
in the CPT experiment \cite{ToganNature2011} (Sec. III).} gives the nuclear
spin dynamics order by order $\dot{p}=(\dot{p})_{1}+(\dot{p})_{2}+\cdots$. The
first-order dynamics describes nuclear spin precession in the electron Knight
fields,%
\begin{equation}
(\dot{p})_{1}=-i\operatorname*{Tr}\nolimits_{e}[\hat{K}+\hat{V}(t),\hat{\rho
}_{0}(t)], \label{EOM1}%
\end{equation}
equivalent to a renormalization of the nuclei Hamiltonian $\hat{H}_{N}$, where
$\hat{\rho}_{0}(t)\equiv\sum_{\mathbf{m},\mathbf{n}}|\mathbf{m}\rangle
\langle\mathbf{n}|p_{\mathbf{m},\mathbf{n}}(t)\hat{P}_{\mathbf{m},\mathbf{n}}$
and $\hat{P}_{\mathbf{m},\mathbf{n}}$ is the normalized electron steady state:
$\mathcal{L}_{\mathbf{m},\mathbf{n}}\hat{P}_{\mathbf{m},\mathbf{n}}=0$. For
$\langle\mathbf{p}|\hat{V}(t)|\mathbf{m}\rangle=\hat{F}e^{-i\omega t}$, we
obtain
\begin{align}
(\dot{p}_{\mathbf{m},\mathbf{m}})_{2}  &  =\sum_{\mathbf{p}}(W_{\mathbf{m}%
\leftarrow\mathbf{p}}p_{\mathbf{p},\mathbf{p}}-W_{\mathbf{p}\leftarrow
\mathbf{m}}p_{\mathbf{m},\mathbf{m}}),\label{EOM2_D}\\
W_{\mathbf{p}\leftarrow\mathbf{m}}  &  =2\operatorname{Re}\int_{0}^{\infty
}e^{i\omega t}\operatorname*{Tr}\nolimits_{e}\hat{F}^{\dagger}e^{\mathcal{L}%
_{\mathbf{p},\mathbf{m}}t}\hat{F}\hat{P}_{\mathbf{m},\mathbf{m}}dt.
\label{WPM}%
\end{align}
The off-diagonal coherence $p_{\mathbf{m},\mathbf{n}}$ ($\mathbf{m}%
\neq\mathbf{n}$) obeys
\begin{align}
(\dot{p}_{\mathbf{m},\mathbf{n}})_{2}  &  =-\left(  \Gamma_{\mathbf{m}%
,\mathbf{n}}^{\varphi}+\frac{\sum_{\mathbf{p}}(W_{\mathbf{p}\leftarrow
\mathbf{n}|\mathbf{m}}+W_{\mathbf{p}\leftarrow\mathbf{m}|\mathbf{n}})}%
{2}\right)  p_{\mathbf{m},\mathbf{n}},\label{EOM2_ND}\\
\Gamma_{\mathbf{m},\mathbf{n}}^{\varphi}  &  \equiv\operatorname{Re}\int
_{0}^{\infty}\operatorname*{Tr}\nolimits_{e}\delta\tilde{K}_{\mathbf{m}%
,\mathbf{n}}e^{\mathcal{L}_{\mathbf{m},\mathbf{n}}t}\delta\tilde
{K}_{\mathbf{m},\mathbf{n}}\hat{P}_{\mathbf{m},\mathbf{n}}dt,
\label{GAMMA_PHI}%
\end{align}
where $\delta\tilde{K}_{\mathbf{m},\mathbf{n}}\equiv\delta\hat{K}%
_{\mathbf{m},\mathbf{n}}-\operatorname*{Tr}_{e}\delta\hat{K}_{\mathbf{m}%
,\mathbf{n}}\hat{P}_{\mathbf{m},\mathbf{n}}$. The expression for
$W_{\mathbf{p}\leftarrow\mathbf{m|n}}$ is involved \cite{Note1}, but it
reduces to $W_{\mathbf{p}\leftarrow\mathbf{m}}$ upon neglecting the difference
between $\hat{K}_{\mathbf{m}}$ and $\hat{K}_{\mathbf{n}}$. The key quantities
of our theory, the transition rate $W_{\mathbf{p}\leftarrow\mathbf{m}}$ [Eq.
(\ref{WPM})] and pure dephasing rate $\Gamma_{\mathbf{m},\mathbf{n}}^{\varphi
}$ [Eq. (\ref{GAMMA_PHI})] are obtained by calculating the inverse
$(\mathcal{L}_{\mathbf{p},\mathbf{m}}+i\omega)^{-1}$ and $\mathcal{L}%
_{\mathbf{m},\mathbf{n}}^{-1}\equiv\lim_{\nu\rightarrow0}(\mathcal{L}%
_{\mathbf{m},\mathbf{n}}+i\nu)^{-1}$. Here we notice that if we focus on the
dynamics of $\{p_{\mathbf{m},\mathbf{m}}\}$ on the time scale $\Delta t\gg
T_{e},T_{2}$, then we can treat $\delta\hat{K}_{\mathbf{m,n}}$ exactly and
still derive Eqs. (\ref{EOM2_D}) and (\ref{WPM}), with $\mathcal{L}%
_{\mathbf{p},\mathbf{m}}$ in Eq. (\ref{WPM}) replaced with $\mathcal{L}%
_{\mathbf{p},\mathbf{m}}^{\mathrm{tot}}\equiv\mathcal{L}_{\mathbf{p}%
,\mathbf{m}}(\bullet)-i\{\bullet,\delta\hat{K}_{\mathbf{p},\mathbf{m}}\}/2$.

Equations (\ref{EOM1}-\ref{GAMMA_PHI}) describe a variety of dissipative and
collective nuclear spin dynamics governed by the NV. Before engineering them,
we present a perturbative expression for $W_{\mathbf{p}\leftarrow\mathbf{m}}$
to exemplify the previously neglected effect of NV coherence. For simplicity
we set $\hat{K}=0$, so $\mathcal{L}_{\mathbf{m},\mathbf{n}}=\mathcal{L}$ and
$\hat{P}_{\mathbf{m},\mathbf{n}}=\hat{P}$ are independent of the nuclei state.
For $\langle\mathbf{p}|\hat{V}(t)|\mathbf{m}\rangle=V_{fi}e^{-i\omega
t}|f\rangle\langle i|$ $(f\neq i)$, $W_{\mathbf{p}\leftarrow\mathbf{m}}$
consists of the golden rule part $2\pi|V_{fi}|^{2}\langle i|\hat{P}%
|i\rangle\delta^{((\Gamma_{f}+\Gamma_{i})/2)}(\operatorname{Re}z_{f,i})$ and
the coherent part $W_{\mathbf{p}\leftarrow\mathbf{m}}^{\mathrm{coh}%
}=2\left\vert V_{f,i}\right\vert ^{2}\operatorname{Im}\sum_{j\neq i}\langle
i|\hat{P}|j\rangle\langle j|\hat{H}_{e}|i\rangle/(z_{f,i}z_{f,j})$, where
$\Gamma_{i}\equiv\sum_{f}\gamma_{fi}$, $\delta^{(\gamma)}(x)\equiv(\gamma
/\pi)/(x^{2}+\gamma^{2})$, and $z_{k,j}\equiv\langle k|\hat{H}_{e}%
|k\rangle-\langle j|\hat{H}_{e}|j\rangle-\omega-i(\Gamma_{k}+\Gamma
_{j}-2\gamma_{k,k}\delta_{k,j})/2$ is the complex energy mismatch. Under
optical/microwave driving, NV coherence and $W_{\mathbf{p}\leftarrow
\mathbf{m}}^{\mathrm{coh}}$ could be important, e.g., it dominates $^{13}$C
nuclei flip by the NV ground state in the first nuclear spin noise suppression
experiment \cite{ToganNature2011} (to be discussed shortly). For more general
$\langle\mathbf{p}|\hat{V}(t)|\mathbf{m}\rangle$, an explicit expression for
$W_{\mathbf{p}\leftarrow\mathbf{m}}$ is given in \cite{Note1}. It can be
easily used to calculate the nuclear spin transition rate in a given
experimental setup. Now we illustrate controlling the nuclear spin evolution
by manipulating the NV.

\textit{Nuclear spin dephasing and relaxation.--}We consider a single $^{13}$C
or $^{14}$N nucleus upon exciting a cyclic NV optical transition
$|g\rangle\leftrightarrow|e\rangle$ with detuning $\Delta$, e.g.,
$|g\rangle=|0\rangle$ and $|e\rangle=|E_{y}\rangle$ in the widely used setup
for single-shot readout \cite{RobledoNature2011,PfaffNatPhys2013}. Dropping
the flip between different NV states, the HFI takes the general form
$\hat{\mathbf{F}}\cdot\hat{\mathbf{I}}$, with $\hat{\mathbf{F}}\equiv
|g\rangle\langle g|\mathbf{a}_{g}+|e\rangle\langle e|\mathbf{a}_{e}$ the NV
Knight field. The NV dissipation includes the radiative decay (rate
$\gamma_{1}$) from $|e\rangle$ to $|g\rangle$ and the pure dephasing (rate
$\gamma_{\varphi}$) of $|e\rangle$. Equation (\ref{EOM1}) gives $(\dot{p}%
)_{1}=-i[\mathbf{\bar{F}}\cdot\hat{\mathbf{I}},\hat{p}]$, where the averaged
Knight field $\mathbf{\bar{F}}\equiv\operatorname*{Tr}\hat{\mathbf{F}}\hat{P}$
defines the nuclear spin $z$-axis $\mathbf{e}_{z}=\mathbf{\bar{F}%
}/|\mathbf{\bar{F}|}$, $\hat{P}$ is the steady NV state with $\langle
e|\hat{P}|e\rangle=W/(\gamma_{1}+2W)$ and $W=2\pi(\Omega_{R}/2)^{2}%
\delta^{((\gamma_{1}+\gamma_{\varphi})/2)}(\Delta)$. For $|\mathbf{a}%
_{g}|,|\mathbf{a}_{e}|\ll\gamma_{1}$, Eqs. (\ref{EOM2_D})-(\ref{GAMMA_PHI})
give the second-order dynamics in the Lindblad form: $(\dot{p})_{2}%
=2\Gamma_{\varphi}\mathcal{D}[\hat{I}_{z}]\hat{p}+\Gamma_{1}(\mathcal{D}%
[\hat{I}_{+}]+\mathcal{D}[\hat{I}_{-}])\hat{p}$, where%
\begin{align}
\Gamma_{\varphi}  &  =c_{0}\langle e|\hat{P}|e\rangle(\gamma_{1}%
+2W)^{-1}(a_{g}^{z}-a_{e}^{z})^{2},\label{DEPHASING}\\
\Gamma_{1}  &  =(c_{0}/2)\langle e|\hat{P}|e\rangle(\gamma_{1}+2W)^{-1}%
|(\mathbf{a}_{g}-\mathbf{a}_{e})_{\perp}|^{2}, \label{RELAXATION}%
\end{align}
and $c_{0}$ is a dimensionless $O(1)$ quantity. The average nuclear spin obeys
$\partial_{t}\langle\hat{I}_{z}\rangle=-\langle\hat{I}_{z}\rangle/T_{1}$ and
$\partial_{t}\langle\hat{I}_{+}\rangle=-\langle\hat{I}_{+}\rangle/T_{2}$ with
$T_{1}=1/(2\Gamma_{1})$ and $T_{2}=1/(\Gamma_{\varphi}+\Gamma_{1})$. Equation
(\ref{DEPHASING}) [(\ref{RELAXATION})] shows that the nuclear spin pure
dephasing (relaxation) is controlled by the fluctuation of the longitudinal
(transverse) Knight field. This was first pointed out in Ref.
\cite{JiangPRL2008}, where a phenomenological spin-fluctuator model was
proposed for numerical simulation. Our analytical results Eqs.
(\ref{DEPHASING}) and (\ref{RELAXATION}) not only provide a microscopic basis
for the previous model \cite{JiangPRL2008} and experimental observations
\cite{DuttScience2007,MaurerScience2012} (e.g., it clearly shows motional
narrowing $\Gamma_{\varphi},\Gamma_{1}\propto1/W$ under saturated pumping
$W\gg\gamma_{1}$), but also demonstrate the possibility \cite{JiangPRL2008} to
control $\Gamma_{\varphi}$ and $\Gamma_{1}$ by a magnetic field $\mathbf{B}%
$:\ since the nuclear Zeeman term $-\gamma_{N}\mathbf{B}\cdot\hat{\mathbf{I}}$
renormalizes $\mathbf{a}_{g/e}$ to $\mathbf{a}_{g/e}-\gamma_{N}\mathbf{B}$, we
can always tune the nuclear spin quantization axis $\mathbf{e}_{z}%
\propto\mathbf{\bar{F}}=\langle g|\hat{P}|g\rangle\mathbf{a}_{g}+\langle
e|\hat{P}|e\rangle\mathbf{a}_{e}-\gamma_{N}\mathbf{B}$ to $\mathbf{e}_{z}%
\perp\mathbf{a}_{g}-\mathbf{a}_{e}$ ($\mathbf{e}_{z}\parallel\mathbf{a}%
_{g}-\mathbf{a}_{e}$) such that $\Gamma_{\varphi}=0$ ($\Gamma_{1}=0$).
Interestingly, the sum rule $\Gamma_{\varphi}+2\Gamma_{1}\propto
|\mathbf{a}_{g}-\mathbf{a}_{e}|^{2}$ suggests that reducing $\Gamma_{\varphi}$
($\Gamma_{1}$) inevitably increases $\Gamma_{1}$ ($\Gamma_{\varphi}$) and it
is impossible to suppress $\Gamma_{\varphi}$ and $\Gamma_{1}$ simultaneously,
unless the NV states are tuned such that $\mathbf{a}_{g}=\mathbf{a}_{e}$. For
more NV levels, analytical results are no longer available, but our general
theory is still applicable.

\textit{Nuclear spin squeezing}.-- Here we explore $^{13}$C nuclear spin
squeezing \cite{RudnerPRL2011} by engineering the first-order evolution Eq.
(\ref{EOM1}). With a magnetic field $B\mathbf{e}_{z}$ to quantize all $^{13}$C
nuclei along $\mathbf{e}_{z}\parallel$ N-V axis and a microwave to couple the
NV ground states $|0\rangle$ and $|+1\rangle$ with detuning $\Delta$, the
rotating frame Hamiltonian consists of $\hat{H}_{e}=(\Omega_{R}/2)(|+1\rangle
\langle0|+h.c.)+\Delta|+1\rangle\langle+1|$ and $\hat{K}\equiv\hat
{h}|+1\rangle\langle+1|$, where $\hat{h}\equiv\sum_{n}\langle+1|\hat
{\mathbf{S}}|+1\rangle\cdot\mathbf{A}_{n}\cdot\hat{\mathbf{I}}_{n}\approx
\sum_{n}a_{n}\hat{I}_{n,z}$ comes from the dipolar HFI with $^{13}$C nuclei
and $a_{n}\equiv\langle+1|\hat{\mathbf{S}}|+1\rangle\cdot\mathbf{A}_{n}%
\cdot\mathbf{e}_{z}$. To introduce \emph{tunable} dissipation, we consider
weak optical pumping of $|+1\rangle$ (with rate $R$) to the orbital excited
state $|e,+1\rangle$, which decays back to $|+1\rangle$ (with rate $\gamma$),
or to a singlet $|S\rangle$ (with rate $\gamma_{\mathrm{ic}}$) and then to
$|0\rangle$ (with rate $\gamma_{s}$). This creates a unidirectional transition
from $|+1\rangle$ to $|0\rangle$. The transition rate $\gamma_{1}$ is tunable
from $R\gamma_{\mathrm{ic}}/(\gamma+R)$ (small $R$) to $\gamma_{s}$ (large
$R$). We define the nuclear spin basis $\{|\mathbf{m}\rangle\}$ as eigenstates
of $\hat{h}$ with eigenvalues $\{h_{\mathbf{m}}\}$. For $\gamma_{1}%
\gg|h_{\mathbf{m}}-h_{\mathbf{n}}|$, Eq. (\ref{EOM1}) gives $(\dot
{p}_{\mathbf{m},\mathbf{n}})_{1}\approx-i\langle\mathbf{m}|[\hat{h}P_{11}%
(\hat{h}),\hat{p}]|\mathbf{n}\rangle$, where $P_{11}(\hat{h})=W(\hat
{h})/(\gamma_{1}+2W(\hat{h}))$ is the $\hat{h}$-dependent population of
$|+1\rangle$ and $W(\hat{h})\equiv2\pi(\Omega_{R}/2)^{2}\delta^{(\gamma
_{1}/2)}(\Delta+\hat{h})$. According to Ref. \cite{RudnerPRL2011}, for
polarized nuclei initially along $\mathbf{e}_{x}$ axis (prepared by rotating
$\mathbf{e}_{z}$-polarized nuclei by a r.f. pulse), the evolution under the
nonlinear Hamiltonian $\hat{h}P_{11}(\hat{h})\approx P_{11}(0)\hat{h}%
+P_{11}^{\prime}(0)\hat{h}^{2}$ could lead to nuclear spin squeezing, even for
non-uniform coupling $\{a_{n}\}$. Taking $a_{n}=a$ for an estimate, the
characteristic squeezing time \cite{KitagawaPRA1993,RudnerPRL2011} for $N$
nuclei is $t_{S}^{(N)}=[P_{11}^{\prime}(0)Na^{2}]^{-1}$. The maximal
$N$-nuclei collective dephasing rate $\sim N^{2}\Gamma_{\varphi}$ with
$\Gamma_{\varphi}$ obtained analogous to Eq. (\ref{DEPHASING}). For
$\Delta=4\Omega_{R}=2\gamma_{1}$, we have $\Gamma_{\varphi}^{(N)}t_{S}%
^{(N)}\approx N/100$, suggesting significant squeezing for $N\ll100$ nuclei
without appreciable dephasing.

\begin{figure}[ptb]
\includegraphics[width=\columnwidth]{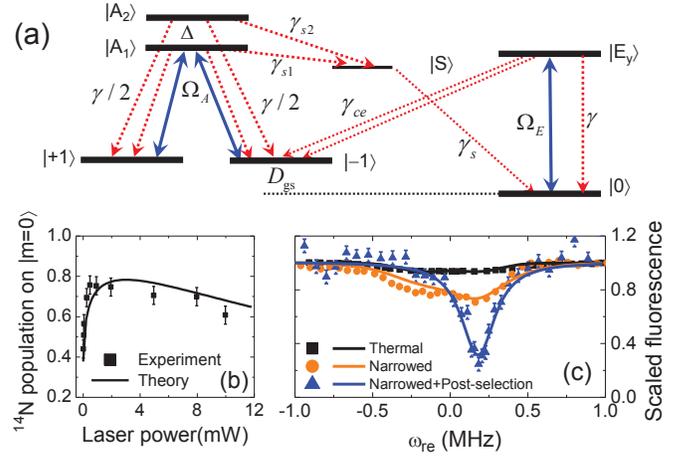} \caption{(color online)
(a) NV center under CPT at low temperature \cite{ToganNature2011}. Solid
(Dashed) arrows denote laser excitation (Lindblad damping). Theoretical
results (lines) vs. experimental data \cite{ToganNature2011} (symbols) for (b)
$^{14}$N population on $|m_{0}=0\rangle$ and (c) NV fluorescence for different
preparation of $^{13}$C states. Relevant parameters: $A_{e}=40\ \mathrm{MHz}$
\cite{FuchsPRL2008}, $A_{g}=2.2\ \mathrm{MHz}$ \cite{DohertyPRB2012},
$\gamma_{s1}=\gamma=1/(12\ \mathrm{ns})$, $\gamma_{s2}=\gamma/120$,
$\gamma_{ce}=\gamma/800$ (obtained by fitting the fluorescence data
\cite{ToganNature2011}), $\gamma_{s}=\gamma/33$ \cite{ToganNature2010}, photon
collection efficiency $\epsilon=5\times10^{-4}$ \cite{ToganNature2011},
$\gamma_{C}=2.5\times10^{-2}\ \mathrm{s}^{-1}$, $T_{\mathrm{cond}%
}=0.288\ \mathrm{ms}$\textbf{,} $\Omega_{A}=2\ \mathrm{MHz}$, readout Rabi
frequency $\Omega_{A}^{\mathrm{re}}=3.2$ (black), $10$ (orange), and
$8\ \mathrm{MHz}$ (blue).}%
\label{G_NVCOOLING}%
\end{figure}

\textit{Controlling nuclear spin noise}.-- Here we consider continuous pumping
the NV to control the noise from many $^{13}$C nuclei coupled to the NV via
dipolar HFI $\sum_{n=1}^{N}\hat{\mathbf{S}}\cdot\mathbf{A}_{n}\cdot
\hat{\mathbf{I}}_{n}$. To focus on noise control, we neglect the squeezing
effect, so Eq. (\ref{EOM1}) gives $(\dot{p})_{1}=-i[\sum_{n}\mathbf{b}%
_{n}\cdot\hat{\mathbf{I}}_{n},\hat{p}]$. Here $\mathbf{b}_{n}\equiv
(\operatorname*{Tr}\mathbf{\hat{S}}\hat{P})\cdot\mathbf{A}_{n}-\gamma
_{N}\mathbf{B}$ defines a local coordinate $(\mathbf{e}_{n,X},\mathbf{e}%
_{n,Y},\mathbf{e}_{n,Z})$ for $\hat{\mathbf{I}}_{n}$, where $\mathbf{e}%
_{n,Z}\propto\mathbf{b}_{n}$ and $\hat{P}$ is the steady NV state in the
absence of $^{13}$C nuclei. We decompose the dipolar HFI into $\hat{K}%
\equiv\hat{\mathbf{S}}\cdot\hat{\mathbf{h}}\approx\hat{S}_{z}\hat{h}_{z}$
($\mathbf{e}_{z}$ along N-V axis) and the remaining part $\hat{V}$, where
$\hat{\mathbf{h}}\equiv\sum_{n}(\mathbf{A}_{n}\cdot\mathbf{e}_{n,Z})\hat
{I}_{n,Z}$. The $\hat{K}$ term not only allows the thermal fluctuation of
$\hat{h}_{z}$ to rapidly decohere the NV spin, but also allows the NV spin
magnetometer to monitor and control the slow $\hat{h}_{z}$ fluctuation by
engineering the following feedback loop \cite{YangPRB2013}. Here for
simplicity we take $\mathbf{A}_{n}=\mathbf{A}$, so that $\hat{h}_{z}=a\sum
_{n}\hat{I}_{n,Z}\leq h_{\max}\equiv aN/2$ and $a\equiv\mathbf{e}_{z}%
\cdot\mathbf{A}_{n}\cdot\mathbf{e}_{n,Z}$, although the physical mechanism is
general: (i) Through $\hat{K}=\hat{S}_{z}\hat{h}_{z}$, the NV spin $\hat
{S}_{z}$ monitors the fluctuation of $\hat{h}_{z}$ and records its
instantaneous value $h$ into the NV steady state $\hat{P}(h)$, as determined
by $-i[\hat{H}_{e}+\hat{S}_{z}h,\hat{P}]+\mathcal{L}_{e}\hat{P}=0$; (ii)
Through $\hat{V}$, the NV state $\hat{P}(h)$ flips each $^{13}$C with rates
$W_{\uparrow\leftarrow\downarrow}(h)$ and $W_{\downarrow\leftarrow\uparrow
}(h)$ [given by Eq. (\ref{EOM2_D}) with $\hat{P}_{\mathbf{m},\mathbf{m}}$
replaced by $\hat{P}(h)$] and drives $\hat{h}_{z}$ from $h$ to the
steady-state value $\mathbb{H}(h)\equiv Na\sum_{n}\langle\hat{I}_{n,Z}\rangle
$, where $\langle\hat{I}_{n,Z}\rangle=(1/2)[W_{\uparrow\leftarrow\downarrow
}(h)-W_{\downarrow\leftarrow\uparrow}(h)]/[W_{\uparrow\leftarrow\downarrow
}(h)+W_{\downarrow\leftarrow\uparrow}(h)]$. For example, to lock $\hat{h}_{z}$
to a pre-defined value $h_{\mathrm{pre}}$, we can design the feedback loop
such that $\mathbb{H}(h_{\mathrm{pre}})=h_{\mathrm{pre}}$ and the derivative
$\mathbb{H}^{\prime}(h_{\mathrm{pre}})<0$, i.e., the NV flips the nuclei to
decrease (increase) $\hat{h}_{z}$ upon detecting $\hat{h}_{z}>h_{\mathrm{pre}%
}$ ($\hat{h}_{z}<h_{\mathrm{pre}}$). To describe this feedback loop, we
quantify the $\hat{h}_{z}$ noise by the width of the $\hat{h}_{z}$
distribution $p(h)\equiv\operatorname*{Tr}\delta(\hat{h}_{z}-h)\hat{p}$ in the
$N$-nuclei steady state $\hat{p}$ and apply the nuclear spin feedback theory
\cite{YangPRB2013} to Eq. (\ref{EOM2_D}) and obtain%
\begin{equation}
p(h)\propto\frac{e^{-(h-h_{\ast})^{2}/(2\sigma^{2})}}{[W_{\uparrow
\leftarrow\downarrow}(h)+W_{\downarrow\leftarrow\uparrow}(h)][1-h\mathbb{H}%
(h)/h_{\max}^{2}]},\label{NOISE}%
\end{equation}
with $h_{\ast}=\mathbb{H}(h_{\ast})$ the most probable value of $\hat{h}_{z}$,
$\sigma^{2}/\sigma_{\mathrm{th}}^{2}\equiv(1-h_{\ast}^{2}/h_{\max}%
^{2})[1-\mathbb{H}^{\prime}(h_{\ast})]^{-1}$, and $\sigma_{\mathrm{th}}%
=a\sqrt{N}/2$ the thermal fluctuation of $\hat{h}_{z}$. Equation (\ref{NOISE})
summarizes three ways to suppress the $\hat{h}_{z}$ noise.\ (i) Narrow
$e^{-(h-h_{\ast})^{2}/(2\sigma^{2})}$ by high nuclear polarization, e.g.,
$h_{\ast}/h_{\max}\approx\pm90\%$ reduces $\sigma$ by a factor of 2. (ii)
Narrow $e^{-(h-h_{\ast})^{2}/(2\sigma^{2})}$ by strong negative feedback
$\mathbb{H}^{\prime}(h_{\ast})\ll-1$. This provides a general,
measurement-free scenario to control the $\hat{h}_{z}$ noise by operating the
NV as a magnetometer: any scheme in which the NV steady state $\hat{P}(h)$ and
hence the NV-induced steady-state nuclear polarization $\mathbb{H}(h)$ is
sensitive to the value $h$ of $\hat{h}_{z}$ could significantly suppress or
amplify the $\hat{h}_{z}$ noise. A possible scheme is to use very weak optical
pumping at the NV ground state anticrossing to polarize $^{13}$C nuclei
without significantly degrading the NV sensitivity $\Delta h\sim$ NV
linewidth, ultimately limited to $\gtrsim1/T_{2,\mathrm{NV}}$ by the
\textit{true} NV dephasing time $T_{2,\mathrm{NV}}$ (not the inhomogeneous
dephasing time $T_{2,\mathrm{NV}}^{\ast}$). For $\Delta h\ll h_{\max}$, we can
tune $h_{\ast}$ to the region with the most negative $\mathbb{H}^{\prime
}(h_{\ast})\sim-h_{\max}/\Delta h$ to reduce $\sigma$ from $\sigma
_{\mathrm{th}}=a\sqrt{N}/2$ to an $N$-independent value $\sigma\sim
\sqrt{a\Delta h}$. For an estimate, we take $h_{\max}=1$ MHz to obtain the
noise reduction factor $\sigma_{\mathrm{th}}/\sigma=\sqrt{h_{\max}/\Delta
h}\approx10$ (for $\Delta h=10\ \mathrm{kHz}$) or $30$ (for $\Delta h=1$ kHz).

The third approach to suppressing $\hat{h}_{z}$ noise is (iii) to generate a
sharp dip in the denominator of Eq. (\ref{NOISE}), e.g., by $\hat{h}_{z}%
$-dependent coherent population trapping (CPT) \cite{IsslerPRL2010}. Now we
show that this mechanism leads to the first observation of $^{13}$C nuclear
spin noise suppression \cite{ToganNature2011}. The setup of Ref.
\cite{ToganNature2011} consists of a $\Lambda$ system ($|\pm1\rangle$ and
$|A_{1}\rangle$) and a two-level system ($|0\rangle$ and $|E_{y}\rangle$),
both under resonant pumping. Relevant processes are shown in Fig.
\ref{G_NVCOOLING}(a) and the NV Hamiltonian can be found in Ref.
\cite{ToganNature2011} or \cite{Note1}. The NV-nuclei coupling includes the
contact HFI $(A_{g}\hat{\mathbf{S}}_{g}+A_{e}\hat{\mathbf{\newline S}}%
_{e})\cdot\hat{\mathbf{I}}_{0}$ with the $^{14}$N nucleus $\hat{\mathbf{I}%
}_{0}$ and the dipolar HFI $\sum_{n=1}^{N}\hat{\mathbf{S}}\cdot\mathbf{A}%
_{n}\cdot\hat{\mathbf{I}}_{n}$ with the $^{13}$C nuclei $\{\hat{\mathbf{I}%
}_{n}\}$, where $\hat{\mathbf{S}}_{g}$ ($\hat{\mathbf{S}}_{e}$) is the NV
ground (excited) state spin and $\hat{\mathbf{S}}\equiv\hat{\mathbf{S}}%
_{g}+\hat{\mathbf{S}}_{e}$. The electron Knight fields on $\hat{\mathbf{I}%
}_{0}$ and $\hat{\mathbf{I}}_{n}$ are along $\mathbf{e}_{z}\parallel$ N-V axis
and $\mathbf{e}_{z}\cdot\mathbf{A}_{n}\equiv a_{n}\mathbf{e}_{n,Z}$,
respectively. So we define the local coordinate $(\mathbf{e}_{n,X}%
,\mathbf{e}_{n,Y},\mathbf{e}_{n,Z})$ and decompose the HFI into $\hat{K}%
=\hat{S}_{g,z}\hat{h}_{z}$ and the remaining part $\hat{V}$, where $\hat
{h}_{z}\equiv A_{g}\hat{I}_{0,z}+\sum_{n}a_{n}\hat{I}_{n,Z}$. We define the
nuclei basis $|\mathbf{m}\rangle\equiv|m_{0}\rangle\otimes_{n=1}^{N}%
|m_{n}\rangle$ as the product of eigenstates of each nucleus: $\hat{I}%
_{0,z}|m_{0}\rangle=m_{0}|m_{0}\rangle$ and $\hat{I}_{n,Z}|m_{n}\rangle
=m_{n}|m_{n}\rangle$. Then we calculate the transition rates from Eq.
(\ref{WPM}) (the $^{13}$C nuclear spin flip by the NV ground state has no
Fermi golden rule part, only the coherent part $W_{\mathbf{p}\leftarrow
\mathbf{m}}^{\mathrm{coh}}$ contributes) and solve Eq. (\ref{EOM2_D})
numerically to obtain the steady-state nuclear spin populations
$\{p_{\mathbf{m},\mathbf{m}}\}$. The intrinsic $^{13}$C-$^{13}$C interaction
and $^{13}$C relaxation is included as a phenomenological depolarization rate
$\gamma_{C}$ for each $^{13}$C nucleus. The calculated $^{14}$N population on
$|m_{0}=0\rangle$ and the relevant time scale $\sim200$ $\mathrm{\mu s}$ (vs.
experimental value $\sim353\pm34$ $\mathrm{\mu s}$) agree reasonably with the
experiment [Fig. \ref{G_NVCOOLING}(b)]. We also confirm that the observed
decrease of the population at large $\Omega_{A}$ arises from the off-resonant
excitation to $|A_{2}\rangle$, as expected in Ref. \cite{ToganNature2011}. An
impressive observation \cite{ToganNature2011} is the suppressed $\hat{h}_{z}$
noise from $^{13}$C nuclei in the absence of appreciable $^{13}$C
polarization, manifested as the narrowed CPT dip of the NV fluorescence. Using
realistic and experimental parameters, we obtain the steady nuclear
populations $\{p_{\mathbf{m},\mathbf{m}}\}$ at the preparation magnetic field
and use them to calculate the unconditional and post-selected population on
$|E_{y}\rangle$ at the readout magnetic field \cite{Note1}. When normalized to
unity at large readout magnetic field $B_{\mathrm{re}}=\omega_{\mathrm{re}%
}/(g_{e}\mu_{B})$, the results agree reasonably with the experimental
fluorescence [Fig. \ref{G_NVCOOLING}(c)].

Finally we use Eq. (\ref{NOISE}) to analyze qualitatively how the NV detects
and suppresses $\hat{h}_{z}$ noise in the CPT experiment
\cite{ToganNature2011}. Given an instantaneous value $h$ of $\hat{h}_{z}$, the
NV rapidly records $h$ as a two-photon detuning in the NV steady-state
$\hat{P}(h)$. Our calculation shows that the NV-induced nuclei flip always
yields vanishing steady-state polarization $\mathbb{H}(h)=0$, so $h_{\ast}=0$,
$\sigma=\sigma_{\mathrm{th}}$, and Eq. (\ref{NOISE}) gives $p(h)\propto
e^{-h^{2}/(2\sigma_{\mathrm{th}}^{2})}[W_{\uparrow\leftarrow\downarrow
}(h)+W_{\downarrow\leftarrow\uparrow}(h)]^{-1}$. The key is that the nuclei
flip rates $W_{\uparrow\leftarrow\downarrow}(h),W_{\downarrow\leftarrow
\uparrow}(h)\propto|E_{y}\rangle$ population \cite{Note1}, which has a sharp
dip at the two-photon resonance $h=0$. This generates a sharp peak in $p(h)$
and hence suppresses the fluctuation of $\hat{h}_{z}$. Further analysis shows
that at $h=0$, off-resonant excitation to $|A_{2}\rangle$ gives rise to
non-vanishing $|E_{y}\rangle$ population and hence $^{13}$C spin flip that
fundamentally limit the noise suppression efficiency. To avoid this
limitation, a possible scheme is to exploit the strain-induced non-vanishing
$\langle E_{y}|\hat{S}_{z}|E_{y}\rangle$ and hence the term $\propto
|E_{y}\rangle\langle E_{y}|(\hat{I}_{+}+\hat{I}_{-})$ of the dipolar HFI. In a
magnetic field that quantizes $^{13}$C nuclei along the N-V axis, a negative
feedback is expected to significantly suppress the $\hat{h}_{z}$ noise without
being limited by off-resonant excitation to $|A_{2}\rangle$.

To summarize, we have presented a quantum theory for using an optically pumped
NV center as a tunable bath to engineer various dissipative and collective
nuclear spin dynamics, as illustrated by the control of the nuclear spin
dephasing, relaxation, and squeezing. It also reveals a general and efficient
way to control the nuclear spin noise and clarifies the physical mechanism
leading to the first observation of nuclear spin noise suppression
\cite{ToganNature2011}. Apart from NV centers, our theory can be readily
applied to other quantum information platforms such as quantum dots and defect
centers to engineer the nuclear spin dynamics for coherence protection
\cite{ZhaoNatNano2011}, quantum feedback control \cite{WisemanPRA1994}, and
preparing non-classical nuclear spin states for quantum metrology
\cite{JonesScience2009} and information storage
\cite{TaylorPRL2003,TaylorPRL2003a}. It could also be extended to explore the
crossover of the nuclei dissipation from Markovian to non-Markovian
\cite{PiiloPRL2008}, and from quantum Zeno to anti-Zeno effect
\cite{KofmanNature2000} with gradually decreasing dissipation of the NV bath
(by reducing the optical pumping strength).

The authors thank Nan Zhao and L. J. Sham for helpful discussions. This work
was supported by NSFC (Grant No. 11274036 and No. 11322542) and the MOST
(Grant No. 2014CB848700).


\begin{thebibliography}{47}
\expandafter\ifx\csname natexlab\endcsname\relax\def\natexlab#1{#1}\fi
\expandafter\ifx\csname bibnamefont\endcsname\relax
  \def\bibnamefont#1{#1}\fi
\expandafter\ifx\csname bibfnamefont\endcsname\relax
  \def\bibfnamefont#1{#1}\fi
\expandafter\ifx\csname citenamefont\endcsname\relax
  \def\citenamefont#1{#1}\fi
\expandafter\ifx\csname url\endcsname\relax
  \def\url#1{\texttt{#1}}\fi
\expandafter\ifx\csname urlprefix\endcsname\relax\def\urlprefix{URL }\fi
\providecommand{\bibinfo}[2]{#2}
\providecommand{\eprint}[2][]{\url{#2}}

\bibitem[{\citenamefont{Gruber et~al.}(1997)\citenamefont{Gruber,
  Dr\"{a}benstedt, Tietz, Fleury, Wrachtrup, and
  Borczyskowski}}]{GruberScience1997}
\bibinfo{author}{\bibfnamefont{A.}~\bibnamefont{Gruber}},
  \bibinfo{author}{\bibfnamefont{A.}~\bibnamefont{Dr\"{a}benstedt}},
  \bibinfo{author}{\bibfnamefont{C.}~\bibnamefont{Tietz}},
  \bibinfo{author}{\bibfnamefont{L.}~\bibnamefont{Fleury}},
  \bibinfo{author}{\bibfnamefont{J.}~\bibnamefont{Wrachtrup}},
  \bibnamefont{and} \bibinfo{author}{\bibfnamefont{C.~v.}
  \bibnamefont{Borczyskowski}}, \bibinfo{journal}{Science}
  \textbf{\bibinfo{volume}{276}}, \bibinfo{pages}{2012} (\bibinfo{year}{1997}).

\bibitem[{\citenamefont{Dutt et~al.}(2007)\citenamefont{Dutt, Childress, Jiang,
  Togan, Maze, Jelezko, Zibrov, Hemmer, and Lukin}}]{DuttScience2007}
\bibinfo{author}{\bibfnamefont{M.~V.~G.} \bibnamefont{Dutt}},
  \bibinfo{author}{\bibfnamefont{L.}~\bibnamefont{Childress}},
  \bibinfo{author}{\bibfnamefont{L.}~\bibnamefont{Jiang}},
  \bibinfo{author}{\bibfnamefont{E.}~\bibnamefont{Togan}},
  \bibinfo{author}{\bibfnamefont{J.}~\bibnamefont{Maze}},
  \bibinfo{author}{\bibfnamefont{F.}~\bibnamefont{Jelezko}},
  \bibinfo{author}{\bibfnamefont{A.~S.} \bibnamefont{Zibrov}},
  \bibinfo{author}{\bibfnamefont{P.~R.} \bibnamefont{Hemmer}},
  \bibnamefont{and} \bibinfo{author}{\bibfnamefont{M.~D.} \bibnamefont{Lukin}},
  \bibinfo{journal}{Science} \textbf{\bibinfo{volume}{316}},
  \bibinfo{pages}{1312} (\bibinfo{year}{2007}).

\bibitem[{\citenamefont{Maze et~al.}(2008)\citenamefont{Maze, Stanwix, Hodges,
  Hong, Taylor, Cappellaro, Jiang, Dutt, Togan, Zibrov
  et~al.}}]{MazeNature2008}
\bibinfo{author}{\bibfnamefont{J.~R.} \bibnamefont{Maze}},
  \bibinfo{author}{\bibfnamefont{P.~L.} \bibnamefont{Stanwix}},
  \bibinfo{author}{\bibfnamefont{J.~S.} \bibnamefont{Hodges}},
  \bibinfo{author}{\bibfnamefont{S.}~\bibnamefont{Hong}},
  \bibinfo{author}{\bibfnamefont{J.~M.} \bibnamefont{Taylor}},
  \bibinfo{author}{\bibfnamefont{P.}~\bibnamefont{Cappellaro}},
  \bibinfo{author}{\bibfnamefont{L.}~\bibnamefont{Jiang}},
  \bibinfo{author}{\bibfnamefont{M.~V.~G.} \bibnamefont{Dutt}},
  \bibinfo{author}{\bibfnamefont{E.}~\bibnamefont{Togan}},
  \bibinfo{author}{\bibfnamefont{A.~S.} \bibnamefont{Zibrov}},
  \bibnamefont{et~al.}, \bibinfo{journal}{Nature}
  \textbf{\bibinfo{volume}{455}}, \bibinfo{pages}{644} (\bibinfo{year}{2008}).

\bibitem[{\citenamefont{Dolde et~al.}(2011)\citenamefont{Dolde, Fedder,
  Doherty, Nobauer, Rempp, Balasubramanian, Wolf, Reinhard, Hollenberg, Jelezko
  et~al.}}]{DoldeNatPhys2011}
\bibinfo{author}{\bibfnamefont{F.}~\bibnamefont{Dolde}},
  \bibinfo{author}{\bibfnamefont{H.}~\bibnamefont{Fedder}},
  \bibinfo{author}{\bibfnamefont{M.~W.} \bibnamefont{Doherty}},
  \bibinfo{author}{\bibfnamefont{T.}~\bibnamefont{Nobauer}},
  \bibinfo{author}{\bibfnamefont{F.}~\bibnamefont{Rempp}},
  \bibinfo{author}{\bibfnamefont{G.}~\bibnamefont{Balasubramanian}},
  \bibinfo{author}{\bibfnamefont{T.}~\bibnamefont{Wolf}},
  \bibinfo{author}{\bibfnamefont{F.}~\bibnamefont{Reinhard}},
  \bibinfo{author}{\bibfnamefont{L.~C.~L.} \bibnamefont{Hollenberg}},
  \bibinfo{author}{\bibfnamefont{F.}~\bibnamefont{Jelezko}},
  \bibnamefont{et~al.}, \bibinfo{journal}{Nat. Phys.}
  \textbf{\bibinfo{volume}{7}}, \bibinfo{pages}{459} (\bibinfo{year}{2011}).

\bibitem[{\citenamefont{Childress et~al.}(2006)\citenamefont{Childress,
  Gurudev~Dutt, Taylor, Zibrov, Jelezko, Wrachtrup, Hemmer, and
  Lukin}}]{ChildressScience2006}
\bibinfo{author}{\bibfnamefont{L.}~\bibnamefont{Childress}},
  \bibinfo{author}{\bibfnamefont{M.~V.} \bibnamefont{Gurudev~Dutt}},
  \bibinfo{author}{\bibfnamefont{J.~M.} \bibnamefont{Taylor}},
  \bibinfo{author}{\bibfnamefont{A.~S.} \bibnamefont{Zibrov}},
  \bibinfo{author}{\bibfnamefont{F.}~\bibnamefont{Jelezko}},
  \bibinfo{author}{\bibfnamefont{J.}~\bibnamefont{Wrachtrup}},
  \bibinfo{author}{\bibfnamefont{P.~R.} \bibnamefont{Hemmer}},
  \bibnamefont{and} \bibinfo{author}{\bibfnamefont{M.~D.} \bibnamefont{Lukin}},
  \bibinfo{journal}{Science} \textbf{\bibinfo{volume}{314}},
  \bibinfo{pages}{281} (\bibinfo{year}{2006}).

\bibitem[{\citenamefont{Togan et~al.}(2010)\citenamefont{Togan, Chu, Trifonov,
  Jiang, Maze, Childress, Dutt, Sorensen, Hemmer, Zibrov
  et~al.}}]{ToganNature2010}
\bibinfo{author}{\bibfnamefont{E.}~\bibnamefont{Togan}},
  \bibinfo{author}{\bibfnamefont{Y.}~\bibnamefont{Chu}},
  \bibinfo{author}{\bibfnamefont{A.~S.} \bibnamefont{Trifonov}},
  \bibinfo{author}{\bibfnamefont{L.}~\bibnamefont{Jiang}},
  \bibinfo{author}{\bibfnamefont{J.}~\bibnamefont{Maze}},
  \bibinfo{author}{\bibfnamefont{L.}~\bibnamefont{Childress}},
  \bibinfo{author}{\bibfnamefont{M.~V.~G.} \bibnamefont{Dutt}},
  \bibinfo{author}{\bibfnamefont{A.~S.} \bibnamefont{Sorensen}},
  \bibinfo{author}{\bibfnamefont{P.~R.} \bibnamefont{Hemmer}},
  \bibinfo{author}{\bibfnamefont{A.~S.} \bibnamefont{Zibrov}},
  \bibnamefont{et~al.}, \bibinfo{journal}{Nature}
  \textbf{\bibinfo{volume}{466}}, \bibinfo{pages}{730} (\bibinfo{year}{2010}).

\bibitem[{\citenamefont{Neumann
  et~al.}(2010{\natexlab{a}})\citenamefont{Neumann, Kolesov, Naydenov, Beck,
  Rempp, Steiner, Jacques, Balasubramanian, Markham, Twitchen
  et~al.}}]{NeumannNatPhys2010}
\bibinfo{author}{\bibfnamefont{P.}~\bibnamefont{Neumann}},
  \bibinfo{author}{\bibfnamefont{R.}~\bibnamefont{Kolesov}},
  \bibinfo{author}{\bibfnamefont{B.}~\bibnamefont{Naydenov}},
  \bibinfo{author}{\bibfnamefont{J.}~\bibnamefont{Beck}},
  \bibinfo{author}{\bibfnamefont{F.}~\bibnamefont{Rempp}},
  \bibinfo{author}{\bibfnamefont{M.}~\bibnamefont{Steiner}},
  \bibinfo{author}{\bibfnamefont{V.}~\bibnamefont{Jacques}},
  \bibinfo{author}{\bibfnamefont{G.}~\bibnamefont{Balasubramanian}},
  \bibinfo{author}{\bibfnamefont{M.~L.} \bibnamefont{Markham}},
  \bibinfo{author}{\bibfnamefont{D.~J.} \bibnamefont{Twitchen}},
  \bibnamefont{et~al.}, \bibinfo{journal}{Nat. Phys.}
  \textbf{\bibinfo{volume}{6}}, \bibinfo{pages}{249}
  (\bibinfo{year}{2010}{\natexlab{a}}).

\bibitem[{\citenamefont{Neumann
  et~al.}(2010{\natexlab{b}})\citenamefont{Neumann, Beck, Steiner, Rempp,
  Fedder, Hemmer, Wrachtrup, and Jelezko}}]{NeumannScience2010}
\bibinfo{author}{\bibfnamefont{P.}~\bibnamefont{Neumann}},
  \bibinfo{author}{\bibfnamefont{J.}~\bibnamefont{Beck}},
  \bibinfo{author}{\bibfnamefont{M.}~\bibnamefont{Steiner}},
  \bibinfo{author}{\bibfnamefont{F.}~\bibnamefont{Rempp}},
  \bibinfo{author}{\bibfnamefont{H.}~\bibnamefont{Fedder}},
  \bibinfo{author}{\bibfnamefont{P.~R.} \bibnamefont{Hemmer}},
  \bibinfo{author}{\bibfnamefont{J.}~\bibnamefont{Wrachtrup}},
  \bibnamefont{and} \bibinfo{author}{\bibfnamefont{F.}~\bibnamefont{Jelezko}},
  \bibinfo{journal}{Science} \textbf{\bibinfo{volume}{329}},
  \bibinfo{pages}{542} (\bibinfo{year}{2010}{\natexlab{b}}).

\bibitem[{\citenamefont{Neumann et~al.}(2008)\citenamefont{Neumann, Mizuochi,
  Rempp, Hemmer, Watanabe, Yamasaki, Jacques, Gaebel, Jelezko, and
  Wrachtrup}}]{NeumannScience2008}
\bibinfo{author}{\bibfnamefont{P.}~\bibnamefont{Neumann}},
  \bibinfo{author}{\bibfnamefont{N.}~\bibnamefont{Mizuochi}},
  \bibinfo{author}{\bibfnamefont{F.}~\bibnamefont{Rempp}},
  \bibinfo{author}{\bibfnamefont{P.}~\bibnamefont{Hemmer}},
  \bibinfo{author}{\bibfnamefont{H.}~\bibnamefont{Watanabe}},
  \bibinfo{author}{\bibfnamefont{S.}~\bibnamefont{Yamasaki}},
  \bibinfo{author}{\bibfnamefont{V.}~\bibnamefont{Jacques}},
  \bibinfo{author}{\bibfnamefont{T.}~\bibnamefont{Gaebel}},
  \bibinfo{author}{\bibfnamefont{F.}~\bibnamefont{Jelezko}}, \bibnamefont{and}
  \bibinfo{author}{\bibfnamefont{J.}~\bibnamefont{Wrachtrup}},
  \bibinfo{journal}{Science} \textbf{\bibinfo{volume}{320}},
  \bibinfo{pages}{1326} (\bibinfo{year}{2008}).

\bibitem[{\citenamefont{Yao et~al.}(2012)\citenamefont{Yao, Jiang, Gorshkov,
  Maurer, Giedke, Cirac, and Lukin}}]{YaoNatCommun2012}
\bibinfo{author}{\bibfnamefont{N.}~\bibnamefont{Yao}},
  \bibinfo{author}{\bibfnamefont{L.}~\bibnamefont{Jiang}},
  \bibinfo{author}{\bibfnamefont{A.}~\bibnamefont{Gorshkov}},
  \bibinfo{author}{\bibfnamefont{P.}~\bibnamefont{Maurer}},
  \bibinfo{author}{\bibfnamefont{G.}~\bibnamefont{Giedke}},
  \bibinfo{author}{\bibfnamefont{J.}~\bibnamefont{Cirac}}, \bibnamefont{and}
  \bibinfo{author}{\bibfnamefont{M.}~\bibnamefont{Lukin}},
  \bibinfo{journal}{Nat. Commun.} \textbf{\bibinfo{volume}{3}},
  \bibinfo{pages}{800} (\bibinfo{year}{2012}).

\bibitem[{\citenamefont{Waldherr et~al.}(2014)\citenamefont{Waldherr, Wang,
  Zaiser, Jamali, Schulte-Herbruggen, Abe, Ohshima, Isoya, Du, Neumann
  et~al.}}]{WaldherrNature2014}
\bibinfo{author}{\bibfnamefont{G.}~\bibnamefont{Waldherr}},
  \bibinfo{author}{\bibfnamefont{Y.}~\bibnamefont{Wang}},
  \bibinfo{author}{\bibfnamefont{S.}~\bibnamefont{Zaiser}},
  \bibinfo{author}{\bibfnamefont{M.}~\bibnamefont{Jamali}},
  \bibinfo{author}{\bibfnamefont{T.}~\bibnamefont{Schulte-Herbruggen}},
  \bibinfo{author}{\bibfnamefont{H.}~\bibnamefont{Abe}},
  \bibinfo{author}{\bibfnamefont{T.}~\bibnamefont{Ohshima}},
  \bibinfo{author}{\bibfnamefont{J.}~\bibnamefont{Isoya}},
  \bibinfo{author}{\bibfnamefont{J.~F.} \bibnamefont{Du}},
  \bibinfo{author}{\bibfnamefont{P.}~\bibnamefont{Neumann}},
  \bibnamefont{et~al.}, \bibinfo{journal}{Nature}
  \textbf{\bibinfo{volume}{506}}, \bibinfo{pages}{204} (\bibinfo{year}{2014}).

\bibitem[{\citenamefont{Taminiau et~al.}(2014)\citenamefont{Taminiau, Cramer,
  van~der Sar, Dobrovitski, and Hanson}}]{TaminiauNatNano2014}
\bibinfo{author}{\bibfnamefont{T.~H.} \bibnamefont{Taminiau}},
  \bibinfo{author}{\bibfnamefont{J.}~\bibnamefont{Cramer}},
  \bibinfo{author}{\bibfnamefont{T.}~\bibnamefont{van~der Sar}},
  \bibinfo{author}{\bibfnamefont{V.~V.} \bibnamefont{Dobrovitski}},
  \bibnamefont{and} \bibinfo{author}{\bibfnamefont{R.}~\bibnamefont{Hanson}},
  \bibinfo{journal}{Nat Nano} \textbf{\bibinfo{volume}{9}},
  \bibinfo{pages}{171} (\bibinfo{year}{2014}).

\bibitem[{\citenamefont{Jiang et~al.}(2008)\citenamefont{Jiang, Dutt, Togan,
  Childress, Cappellaro, Taylor, and Lukin}}]{JiangPRL2008}
\bibinfo{author}{\bibfnamefont{L.}~\bibnamefont{Jiang}},
  \bibinfo{author}{\bibfnamefont{M.~V.~G.} \bibnamefont{Dutt}},
  \bibinfo{author}{\bibfnamefont{E.}~\bibnamefont{Togan}},
  \bibinfo{author}{\bibfnamefont{L.}~\bibnamefont{Childress}},
  \bibinfo{author}{\bibfnamefont{P.}~\bibnamefont{Cappellaro}},
  \bibinfo{author}{\bibfnamefont{J.~M.} \bibnamefont{Taylor}},
  \bibnamefont{and} \bibinfo{author}{\bibfnamefont{M.~D.} \bibnamefont{Lukin}},
  \bibinfo{journal}{Phys. Rev. Lett.} \textbf{\bibinfo{volume}{100}},
  \bibinfo{pages}{073001} (\bibinfo{year}{2008}).

\bibitem[{\citenamefont{Dreau et~al.}(2013)\citenamefont{Dreau, Spinicelli,
  Maze, Roch, and Jacques}}]{DreauPRL2013}
\bibinfo{author}{\bibfnamefont{A.}~\bibnamefont{Dreau}},
  \bibinfo{author}{\bibfnamefont{P.}~\bibnamefont{Spinicelli}},
  \bibinfo{author}{\bibfnamefont{J.~R.} \bibnamefont{Maze}},
  \bibinfo{author}{\bibfnamefont{J.-F.} \bibnamefont{Roch}}, \bibnamefont{and}
  \bibinfo{author}{\bibfnamefont{V.}~\bibnamefont{Jacques}},
  \bibinfo{journal}{Phys. Rev. Lett.} \textbf{\bibinfo{volume}{110}},
  \bibinfo{pages}{060502} (\bibinfo{year}{2013}).

\bibitem[{\citenamefont{King et~al.}(2010)\citenamefont{King, Coles, and
  Reimer}}]{KingPRB2010}
\bibinfo{author}{\bibfnamefont{J.~P.} \bibnamefont{King}},
  \bibinfo{author}{\bibfnamefont{P.~J.} \bibnamefont{Coles}}, \bibnamefont{and}
  \bibinfo{author}{\bibfnamefont{J.~A.} \bibnamefont{Reimer}},
  \bibinfo{journal}{Phys. Rev. B} \textbf{\bibinfo{volume}{81}},
  \bibinfo{pages}{073201} (\bibinfo{year}{2010}).

\bibitem[{\citenamefont{Fischer
  et~al.}(2013{\natexlab{a}})\citenamefont{Fischer, Bretschneider, London,
  Budker, Gershoni, and Frydman}}]{FischerPRL2013}
\bibinfo{author}{\bibfnamefont{R.}~\bibnamefont{Fischer}},
  \bibinfo{author}{\bibfnamefont{C.~O.} \bibnamefont{Bretschneider}},
  \bibinfo{author}{\bibfnamefont{P.}~\bibnamefont{London}},
  \bibinfo{author}{\bibfnamefont{D.}~\bibnamefont{Budker}},
  \bibinfo{author}{\bibfnamefont{D.}~\bibnamefont{Gershoni}}, \bibnamefont{and}
  \bibinfo{author}{\bibfnamefont{L.}~\bibnamefont{Frydman}},
  \bibinfo{journal}{Phys. Rev. Lett.} \textbf{\bibinfo{volume}{111}},
  \bibinfo{pages}{057601} (\bibinfo{year}{2013}{\natexlab{a}}).

\bibitem[{\citenamefont{Fischer
  et~al.}(2013{\natexlab{b}})\citenamefont{Fischer, Jarmola, Kehayias, and
  Budker}}]{FischerPRB2013}
\bibinfo{author}{\bibfnamefont{R.}~\bibnamefont{Fischer}},
  \bibinfo{author}{\bibfnamefont{A.}~\bibnamefont{Jarmola}},
  \bibinfo{author}{\bibfnamefont{P.}~\bibnamefont{Kehayias}}, \bibnamefont{and}
  \bibinfo{author}{\bibfnamefont{D.}~\bibnamefont{Budker}},
  \bibinfo{journal}{Phys. Rev. B} \textbf{\bibinfo{volume}{87}},
  \bibinfo{pages}{125207} (\bibinfo{year}{2013}{\natexlab{b}}).

\bibitem[{\citenamefont{Wang et~al.}(2013)\citenamefont{Wang, Shin, Avalos,
  Seltzer, Budker, Pines, and Bajaj}}]{WangNatCommun2013}
\bibinfo{author}{\bibfnamefont{H.-J.} \bibnamefont{Wang}},
  \bibinfo{author}{\bibfnamefont{C.~S.} \bibnamefont{Shin}},
  \bibinfo{author}{\bibfnamefont{C.~E.} \bibnamefont{Avalos}},
  \bibinfo{author}{\bibfnamefont{S.~J.} \bibnamefont{Seltzer}},
  \bibinfo{author}{\bibfnamefont{D.}~\bibnamefont{Budker}},
  \bibinfo{author}{\bibfnamefont{A.}~\bibnamefont{Pines}}, \bibnamefont{and}
  \bibinfo{author}{\bibfnamefont{V.~S.} \bibnamefont{Bajaj}},
  \bibinfo{journal}{Nat. Commun.} \textbf{\bibinfo{volume}{4}},
  \bibinfo{pages}{1} (\bibinfo{year}{2013}).

\bibitem[{\citenamefont{Togan et~al.}(2011)\citenamefont{Togan, Chu, Imamoglu,
  and Lukin}}]{ToganNature2011}
\bibinfo{author}{\bibfnamefont{E.}~\bibnamefont{Togan}},
  \bibinfo{author}{\bibfnamefont{Y.}~\bibnamefont{Chu}},
  \bibinfo{author}{\bibfnamefont{A.}~\bibnamefont{Imamoglu}}, \bibnamefont{and}
  \bibinfo{author}{\bibfnamefont{M.~D.} \bibnamefont{Lukin}},
  \bibinfo{journal}{Nature} \textbf{\bibinfo{volume}{478}},
  \bibinfo{pages}{497} (\bibinfo{year}{2011}).

\bibitem[{\citenamefont{Balasubramanian
  et~al.}(2008)\citenamefont{Balasubramanian, Chan, Kolesov, Al-Hmoud, Tisler,
  Shin, Kim, Wojcik, Hemmer, Krueger et~al.}}]{BalasubramanianNature2008}
\bibinfo{author}{\bibfnamefont{G.}~\bibnamefont{Balasubramanian}},
  \bibinfo{author}{\bibfnamefont{I.~Y.} \bibnamefont{Chan}},
  \bibinfo{author}{\bibfnamefont{R.}~\bibnamefont{Kolesov}},
  \bibinfo{author}{\bibfnamefont{M.}~\bibnamefont{Al-Hmoud}},
  \bibinfo{author}{\bibfnamefont{J.}~\bibnamefont{Tisler}},
  \bibinfo{author}{\bibfnamefont{C.}~\bibnamefont{Shin}},
  \bibinfo{author}{\bibfnamefont{C.}~\bibnamefont{Kim}},
  \bibinfo{author}{\bibfnamefont{A.}~\bibnamefont{Wojcik}},
  \bibinfo{author}{\bibfnamefont{P.~R.} \bibnamefont{Hemmer}},
  \bibinfo{author}{\bibfnamefont{A.}~\bibnamefont{Krueger}},
  \bibnamefont{et~al.}, \bibinfo{journal}{Nature}
  \textbf{\bibinfo{volume}{455}}, \bibinfo{pages}{648} (\bibinfo{year}{2008}).

\bibitem[{\citenamefont{Jones et~al.}(2009)\citenamefont{Jones, Karlen,
  Fitzsimons, Ardavan, Benjamin, Briggs, and Morton}}]{JonesScience2009}
\bibinfo{author}{\bibfnamefont{J.~A.} \bibnamefont{Jones}},
  \bibinfo{author}{\bibfnamefont{S.~D.} \bibnamefont{Karlen}},
  \bibinfo{author}{\bibfnamefont{J.}~\bibnamefont{Fitzsimons}},
  \bibinfo{author}{\bibfnamefont{A.}~\bibnamefont{Ardavan}},
  \bibinfo{author}{\bibfnamefont{S.~C.} \bibnamefont{Benjamin}},
  \bibinfo{author}{\bibfnamefont{G.~A.~D.} \bibnamefont{Briggs}},
  \bibnamefont{and} \bibinfo{author}{\bibfnamefont{J.~J.~L.}
  \bibnamefont{Morton}}, \bibinfo{journal}{Science}
  \textbf{\bibinfo{volume}{324}}, \bibinfo{pages}{1166} (\bibinfo{year}{2009}).

\bibitem[{\citenamefont{Waldherr et~al.}(2012)\citenamefont{Waldherr, Beck,
  Neumann, Said, Nitsche, Markham, Twitchen, Twamley, Jelezko, and
  Wrachtrup}}]{WaldherrNatNano2012}
\bibinfo{author}{\bibfnamefont{G.}~\bibnamefont{Waldherr}},
  \bibinfo{author}{\bibfnamefont{J.}~\bibnamefont{Beck}},
  \bibinfo{author}{\bibfnamefont{P.}~\bibnamefont{Neumann}},
  \bibinfo{author}{\bibfnamefont{R.~S.} \bibnamefont{Said}},
  \bibinfo{author}{\bibfnamefont{M.}~\bibnamefont{Nitsche}},
  \bibinfo{author}{\bibfnamefont{M.~L.} \bibnamefont{Markham}},
  \bibinfo{author}{\bibfnamefont{D.~J.} \bibnamefont{Twitchen}},
  \bibinfo{author}{\bibfnamefont{J.}~\bibnamefont{Twamley}},
  \bibinfo{author}{\bibfnamefont{F.}~\bibnamefont{Jelezko}}, \bibnamefont{and}
  \bibinfo{author}{\bibfnamefont{J.}~\bibnamefont{Wrachtrup}},
  \bibinfo{journal}{Nat. Nanotechnol.} \textbf{\bibinfo{volume}{7}},
  \bibinfo{pages}{105} (\bibinfo{year}{2012}).

\bibitem[{\citenamefont{Taylor et~al.}(2003{\natexlab{a}})\citenamefont{Taylor,
  Imamoglu, and Lukin}}]{TaylorPRL2003}
\bibinfo{author}{\bibfnamefont{J.~M.} \bibnamefont{Taylor}},
  \bibinfo{author}{\bibfnamefont{A.}~\bibnamefont{Imamoglu}}, \bibnamefont{and}
  \bibinfo{author}{\bibfnamefont{M.~D.} \bibnamefont{Lukin}},
  \bibinfo{journal}{Phys. Rev. Lett.} \textbf{\bibinfo{volume}{91}},
  \bibinfo{pages}{246802} (\bibinfo{year}{2003}{\natexlab{a}}).

\bibitem[{\citenamefont{Taylor et~al.}(2003{\natexlab{b}})\citenamefont{Taylor,
  Marcus, and Lukin}}]{TaylorPRL2003a}
\bibinfo{author}{\bibfnamefont{J.~M.} \bibnamefont{Taylor}},
  \bibinfo{author}{\bibfnamefont{C.~M.} \bibnamefont{Marcus}},
  \bibnamefont{and} \bibinfo{author}{\bibfnamefont{M.~D.} \bibnamefont{Lukin}},
  \bibinfo{journal}{Phys. Rev. Lett.} \textbf{\bibinfo{volume}{90}},
  \bibinfo{pages}{206803} (\bibinfo{year}{2003}{\natexlab{b}}).

\bibitem[{\citenamefont{Jacques et~al.}(2009)\citenamefont{Jacques, Neumann,
  Beck, Markham, Twitchen, Meijer, Kaiser, Balasubramanian, Jelezko, and
  Wrachtrup}}]{JacquesPRL2009}
\bibinfo{author}{\bibfnamefont{V.}~\bibnamefont{Jacques}},
  \bibinfo{author}{\bibfnamefont{P.}~\bibnamefont{Neumann}},
  \bibinfo{author}{\bibfnamefont{J.}~\bibnamefont{Beck}},
  \bibinfo{author}{\bibfnamefont{M.}~\bibnamefont{Markham}},
  \bibinfo{author}{\bibfnamefont{D.}~\bibnamefont{Twitchen}},
  \bibinfo{author}{\bibfnamefont{J.}~\bibnamefont{Meijer}},
  \bibinfo{author}{\bibfnamefont{F.}~\bibnamefont{Kaiser}},
  \bibinfo{author}{\bibfnamefont{G.}~\bibnamefont{Balasubramanian}},
  \bibinfo{author}{\bibfnamefont{F.}~\bibnamefont{Jelezko}}, \bibnamefont{and}
  \bibinfo{author}{\bibfnamefont{J.}~\bibnamefont{Wrachtrup}},
  \bibinfo{journal}{Phys. Rev. Lett.} \textbf{\bibinfo{volume}{102}},
  \bibinfo{pages}{057403} (\bibinfo{year}{2009}).

\bibitem[{\citenamefont{London et~al.}(2013)\citenamefont{London, Scheuer, Cai,
  Schwarz, Retzker, Plenio, Katagiri, Teraji, Koizumi, Isoya
  et~al.}}]{LondonPRL2013}
\bibinfo{author}{\bibfnamefont{P.}~\bibnamefont{London}},
  \bibinfo{author}{\bibfnamefont{J.}~\bibnamefont{Scheuer}},
  \bibinfo{author}{\bibfnamefont{J.-M.} \bibnamefont{Cai}},
  \bibinfo{author}{\bibfnamefont{I.}~\bibnamefont{Schwarz}},
  \bibinfo{author}{\bibfnamefont{A.}~\bibnamefont{Retzker}},
  \bibinfo{author}{\bibfnamefont{M.~B.} \bibnamefont{Plenio}},
  \bibinfo{author}{\bibfnamefont{M.}~\bibnamefont{Katagiri}},
  \bibinfo{author}{\bibfnamefont{T.}~\bibnamefont{Teraji}},
  \bibinfo{author}{\bibfnamefont{S.}~\bibnamefont{Koizumi}},
  \bibinfo{author}{\bibfnamefont{J.}~\bibnamefont{Isoya}},
  \bibnamefont{et~al.}, \bibinfo{journal}{Phys. Rev. Lett.}
  \textbf{\bibinfo{volume}{111}}, \bibinfo{pages}{067601}
  (\bibinfo{year}{2013}).

\bibitem[{\citenamefont{Liu et~al.}(2014)\citenamefont{Liu, Jiang, Chang, Liu,
  Li, Gu, Po, Zhang, Zhao, and Pan}}]{LiuNanoscale2014}
\bibinfo{author}{\bibfnamefont{G.-Q.} \bibnamefont{Liu}},
  \bibinfo{author}{\bibfnamefont{Q.-Q.} \bibnamefont{Jiang}},
  \bibinfo{author}{\bibfnamefont{Y.-C.} \bibnamefont{Chang}},
  \bibinfo{author}{\bibfnamefont{D.-Q.} \bibnamefont{Liu}},
  \bibinfo{author}{\bibfnamefont{W.-X.} \bibnamefont{Li}},
  \bibinfo{author}{\bibfnamefont{C.-Z.} \bibnamefont{Gu}},
  \bibinfo{author}{\bibfnamefont{H.~C.} \bibnamefont{Po}},
  \bibinfo{author}{\bibfnamefont{W.-X.} \bibnamefont{Zhang}},
  \bibinfo{author}{\bibfnamefont{N.}~\bibnamefont{Zhao}}, \bibnamefont{and}
  \bibinfo{author}{\bibfnamefont{X.-Y.} \bibnamefont{Pan}},
  \bibinfo{journal}{Nanoscale} \textbf{\bibinfo{volume}{6}},
  \bibinfo{pages}{10134} (\bibinfo{year}{2014}).

\bibitem[{\citenamefont{Dr\'eau et~al.}(2014)\citenamefont{Dr\'eau, Jamonneau,
  Gazzano, Kosen, Roch, Maze, and Jacques}}]{DreauPRL2014}
\bibinfo{author}{\bibfnamefont{A.}~\bibnamefont{Dr\'eau}},
  \bibinfo{author}{\bibfnamefont{P.}~\bibnamefont{Jamonneau}},
  \bibinfo{author}{\bibfnamefont{O.}~\bibnamefont{Gazzano}},
  \bibinfo{author}{\bibfnamefont{S.}~\bibnamefont{Kosen}},
  \bibinfo{author}{\bibfnamefont{J.-F.} \bibnamefont{Roch}},
  \bibinfo{author}{\bibfnamefont{J.~R.} \bibnamefont{Maze}}, \bibnamefont{and}
  \bibinfo{author}{\bibfnamefont{V.}~\bibnamefont{Jacques}},
  \bibinfo{journal}{Phys. Rev. Lett.} \textbf{\bibinfo{volume}{113}},
  \bibinfo{pages}{137601} (\bibinfo{year}{2014}).

\bibitem[{\citenamefont{Pagliero et~al.}(2014)\citenamefont{Pagliero, Laraoui,
  Henshaw, and Meriles}}]{PaglieroAPL2014}
\bibinfo{author}{\bibfnamefont{D.}~\bibnamefont{Pagliero}},
  \bibinfo{author}{\bibfnamefont{A.}~\bibnamefont{Laraoui}},
  \bibinfo{author}{\bibfnamefont{J.~D.} \bibnamefont{Henshaw}},
  \bibnamefont{and} \bibinfo{author}{\bibfnamefont{C.~A.}
  \bibnamefont{Meriles}}, \bibinfo{journal}{Appl. Phys. Lett.}
  \textbf{\bibinfo{volume}{105}}, \bibinfo{eid}{242402} (\bibinfo{year}{2014}).

\bibitem[{\citenamefont{Alvarez et~al.}(2014)\citenamefont{Alvarez,
  Bretschneider, Fischer, London, Kanda, Onoda, Isoya, Gershoni, and
  Frydman}}]{AlvarezArxiv2014}
\bibinfo{author}{\bibfnamefont{G.}~\bibnamefont{Alvarez}},
  \bibinfo{author}{\bibfnamefont{C.}~\bibnamefont{Bretschneider}},
  \bibinfo{author}{\bibfnamefont{R.}~\bibnamefont{Fischer}},
  \bibinfo{author}{\bibfnamefont{P.}~\bibnamefont{London}},
  \bibinfo{author}{\bibfnamefont{H.}~\bibnamefont{Kanda}},
  \bibinfo{author}{\bibfnamefont{S.}~\bibnamefont{Onoda}},
  \bibinfo{author}{\bibfnamefont{J.}~\bibnamefont{Isoya}},
  \bibinfo{author}{\bibfnamefont{D.}~\bibnamefont{Gershoni}}, \bibnamefont{and}
  \bibinfo{author}{\bibfnamefont{L.}~\bibnamefont{Frydman}},
  \bibinfo{journal}{arXiv.} \textbf{\bibinfo{volume}{1412}},
  \bibinfo{pages}{8635} (\bibinfo{year}{2014}).

\bibitem[{\citenamefont{Maurer et~al.}(2012)\citenamefont{Maurer, Kucsko,
  Latta, Jiang, Yao, Bennett, Pastawski, Hunger, Chisholm, Markham
  et~al.}}]{MaurerScience2012}
\bibinfo{author}{\bibfnamefont{P.~C.} \bibnamefont{Maurer}},
  \bibinfo{author}{\bibfnamefont{G.}~\bibnamefont{Kucsko}},
  \bibinfo{author}{\bibfnamefont{C.}~\bibnamefont{Latta}},
  \bibinfo{author}{\bibfnamefont{L.}~\bibnamefont{Jiang}},
  \bibinfo{author}{\bibfnamefont{N.~Y.} \bibnamefont{Yao}},
  \bibinfo{author}{\bibfnamefont{S.~D.} \bibnamefont{Bennett}},
  \bibinfo{author}{\bibfnamefont{F.}~\bibnamefont{Pastawski}},
  \bibinfo{author}{\bibfnamefont{D.}~\bibnamefont{Hunger}},
  \bibinfo{author}{\bibfnamefont{N.}~\bibnamefont{Chisholm}},
  \bibinfo{author}{\bibfnamefont{M.}~\bibnamefont{Markham}},
  \bibnamefont{et~al.}, \bibinfo{journal}{Science}
  \textbf{\bibinfo{volume}{336}}, \bibinfo{pages}{1283} (\bibinfo{year}{2012}).

\bibitem[{\citenamefont{Cirac et~al.}(1992)\citenamefont{Cirac, Blatt, Zoller,
  and Phillips}}]{CiracPRA1992}
\bibinfo{author}{\bibfnamefont{J.~I.} \bibnamefont{Cirac}},
  \bibinfo{author}{\bibfnamefont{R.}~\bibnamefont{Blatt}},
  \bibinfo{author}{\bibfnamefont{P.}~\bibnamefont{Zoller}}, \bibnamefont{and}
  \bibinfo{author}{\bibfnamefont{W.~D.} \bibnamefont{Phillips}},
  \bibinfo{journal}{Phys. Rev. A} \textbf{\bibinfo{volume}{46}},
  \bibinfo{pages}{2668} (\bibinfo{year}{1992}).

\bibitem[{\citenamefont{Wiseman and Milburn}(1993)}]{WisemanPRA1993a}
\bibinfo{author}{\bibfnamefont{H.~M.} \bibnamefont{Wiseman}} \bibnamefont{and}
  \bibinfo{author}{\bibfnamefont{G.~J.} \bibnamefont{Milburn}},
  \bibinfo{journal}{Phys. Rev. A} \textbf{\bibinfo{volume}{47}},
  \bibinfo{pages}{642} (\bibinfo{year}{1993}).

\bibitem[{\citenamefont{Yang and Sham}(2012)}]{YangPRB2012}
\bibinfo{author}{\bibfnamefont{W.}~\bibnamefont{Yang}} \bibnamefont{and}
  \bibinfo{author}{\bibfnamefont{L.~J.} \bibnamefont{Sham}},
  \bibinfo{journal}{Phys. Rev. B} \textbf{\bibinfo{volume}{85}},
  \bibinfo{pages}{235319} (\bibinfo{year}{2012}).

\bibitem[{Note1()}]{Note1}
Note1, \bibinfo{note}{see supplementary material for derivation of Eqs.
  (2)--(6) (Sec. I), a perturbative, explicit expression for $W_{\protect
  \mathbf {p}\leftarrow \protect \mathbf {m}}$ (Sec. II), and summary of the NV
  Hamiltonian, NV-induced nuclear spin transition rates, and calculation of NV
  fluorescence in the CPT experiment \cite {ToganNature2011} (Sec. III).}

\bibitem[{\citenamefont{Robledo et~al.}(2011)\citenamefont{Robledo, Childress,
  Bernien, Hensen, Alkemade, and Hanson}}]{RobledoNature2011}
\bibinfo{author}{\bibfnamefont{L.}~\bibnamefont{Robledo}},
  \bibinfo{author}{\bibfnamefont{L.}~\bibnamefont{Childress}},
  \bibinfo{author}{\bibfnamefont{H.}~\bibnamefont{Bernien}},
  \bibinfo{author}{\bibfnamefont{B.}~\bibnamefont{Hensen}},
  \bibinfo{author}{\bibfnamefont{P.~F.~A.} \bibnamefont{Alkemade}},
  \bibnamefont{and} \bibinfo{author}{\bibfnamefont{R.}~\bibnamefont{Hanson}},
  \bibinfo{journal}{Nature} \textbf{\bibinfo{volume}{477}},
  \bibinfo{pages}{574} (\bibinfo{year}{2011}).

\bibitem[{\citenamefont{Pfaff et~al.}(2013)\citenamefont{Pfaff, Taminiau,
  Robledo, Bernien, Markham, Twitchen, and Hanson}}]{PfaffNatPhys2013}
\bibinfo{author}{\bibfnamefont{W.}~\bibnamefont{Pfaff}},
  \bibinfo{author}{\bibfnamefont{T.~H.} \bibnamefont{Taminiau}},
  \bibinfo{author}{\bibfnamefont{L.}~\bibnamefont{Robledo}},
  \bibinfo{author}{\bibfnamefont{H.}~\bibnamefont{Bernien}},
  \bibinfo{author}{\bibfnamefont{M.}~\bibnamefont{Markham}},
  \bibinfo{author}{\bibfnamefont{D.~J.} \bibnamefont{Twitchen}},
  \bibnamefont{and} \bibinfo{author}{\bibfnamefont{R.}~\bibnamefont{Hanson}},
  \bibinfo{journal}{Nat. Phys.} \textbf{\bibinfo{volume}{9}},
  \bibinfo{pages}{29} (\bibinfo{year}{2013}).

\bibitem[{\citenamefont{Rudner et~al.}(2011)\citenamefont{Rudner, Vandersypen,
  Vuleti\ifmmode~\acute{c}\else \'{c}\fi{}, and Levitov}}]{RudnerPRL2011}
\bibinfo{author}{\bibfnamefont{M.~S.} \bibnamefont{Rudner}},
  \bibinfo{author}{\bibfnamefont{L.~M.~K.} \bibnamefont{Vandersypen}},
  \bibinfo{author}{\bibfnamefont{V.}~\bibnamefont{Vuleti\ifmmode~\acute{c}\else
  \'{c}\fi{}}}, \bibnamefont{and} \bibinfo{author}{\bibfnamefont{L.~S.}
  \bibnamefont{Levitov}}, \bibinfo{journal}{Phys. Rev. Lett.}
  \textbf{\bibinfo{volume}{107}}, \bibinfo{pages}{206806}
  (\bibinfo{year}{2011}).

\bibitem[{\citenamefont{Kitagawa and Ueda}(1993)}]{KitagawaPRA1993}
\bibinfo{author}{\bibfnamefont{M.}~\bibnamefont{Kitagawa}} \bibnamefont{and}
  \bibinfo{author}{\bibfnamefont{M.}~\bibnamefont{Ueda}},
  \bibinfo{journal}{Phys. Rev. A} \textbf{\bibinfo{volume}{47}},
  \bibinfo{pages}{5138} (\bibinfo{year}{1993}).

\bibitem[{\citenamefont{Fuchs et~al.}(2008)\citenamefont{Fuchs, Dobrovitski,
  Hanson, Batra, Weis, Schenkel, and Awschalom}}]{FuchsPRL2008}
\bibinfo{author}{\bibfnamefont{G.~D.} \bibnamefont{Fuchs}},
  \bibinfo{author}{\bibfnamefont{V.~V.} \bibnamefont{Dobrovitski}},
  \bibinfo{author}{\bibfnamefont{R.}~\bibnamefont{Hanson}},
  \bibinfo{author}{\bibfnamefont{A.}~\bibnamefont{Batra}},
  \bibinfo{author}{\bibfnamefont{C.~D.} \bibnamefont{Weis}},
  \bibinfo{author}{\bibfnamefont{T.}~\bibnamefont{Schenkel}}, \bibnamefont{and}
  \bibinfo{author}{\bibfnamefont{D.~D.} \bibnamefont{Awschalom}},
  \bibinfo{journal}{Phys. Rev. Lett.} \textbf{\bibinfo{volume}{101}},
  \bibinfo{pages}{117601} (\bibinfo{year}{2008}).

\bibitem[{\citenamefont{Doherty et~al.}(2012)\citenamefont{Doherty, Dolde,
  Fedder, Jelezko, Wrachtrup, Manson, and Hollenberg}}]{DohertyPRB2012}
\bibinfo{author}{\bibfnamefont{M.~W.} \bibnamefont{Doherty}},
  \bibinfo{author}{\bibfnamefont{F.}~\bibnamefont{Dolde}},
  \bibinfo{author}{\bibfnamefont{H.}~\bibnamefont{Fedder}},
  \bibinfo{author}{\bibfnamefont{F.}~\bibnamefont{Jelezko}},
  \bibinfo{author}{\bibfnamefont{J.}~\bibnamefont{Wrachtrup}},
  \bibinfo{author}{\bibfnamefont{N.~B.} \bibnamefont{Manson}},
  \bibnamefont{and} \bibinfo{author}{\bibfnamefont{L.~C.~L.}
  \bibnamefont{Hollenberg}}, \bibinfo{journal}{Phys. Rev. B}
  \textbf{\bibinfo{volume}{85}}, \bibinfo{pages}{205203}
  (\bibinfo{year}{2012}).

\bibitem[{\citenamefont{Yang and Sham}(2013)}]{YangPRB2013}
\bibinfo{author}{\bibfnamefont{W.}~\bibnamefont{Yang}} \bibnamefont{and}
  \bibinfo{author}{\bibfnamefont{L.~J.} \bibnamefont{Sham}},
  \bibinfo{journal}{Phys. Rev. B} \textbf{\bibinfo{volume}{88}},
  \bibinfo{pages}{235304} (\bibinfo{year}{2013}).

\bibitem[{\citenamefont{Issler et~al.}(2010)\citenamefont{Issler, Kessler,
  Giedke, Yelin, Cirac, Lukin, and Imamoglu}}]{IsslerPRL2010}
\bibinfo{author}{\bibfnamefont{M.}~\bibnamefont{Issler}},
  \bibinfo{author}{\bibfnamefont{E.~M.} \bibnamefont{Kessler}},
  \bibinfo{author}{\bibfnamefont{G.}~\bibnamefont{Giedke}},
  \bibinfo{author}{\bibfnamefont{S.}~\bibnamefont{Yelin}},
  \bibinfo{author}{\bibfnamefont{I.}~\bibnamefont{Cirac}},
  \bibinfo{author}{\bibfnamefont{M.~D.} \bibnamefont{Lukin}}, \bibnamefont{and}
  \bibinfo{author}{\bibfnamefont{A.}~\bibnamefont{Imamoglu}},
  \bibinfo{journal}{Phys. Rev. Lett.} \textbf{\bibinfo{volume}{105}},
  \bibinfo{pages}{267202} (\bibinfo{year}{2010}).

\bibitem[{\citenamefont{Zhao et~al.}(2011)\citenamefont{Zhao, Hu, Ho, Wan, and
  B.}}]{ZhaoNatNano2011}
\bibinfo{author}{\bibfnamefont{N.}~\bibnamefont{Zhao}},
  \bibinfo{author}{\bibfnamefont{J.-L.} \bibnamefont{Hu}},
  \bibinfo{author}{\bibfnamefont{S.-W.} \bibnamefont{Ho}},
  \bibinfo{author}{\bibfnamefont{J.~T.~K.} \bibnamefont{Wan}},
  \bibnamefont{and} \bibinfo{author}{\bibfnamefont{L.}~\bibnamefont{B.}},
  \bibinfo{journal}{Nat. Nanotechnol.} \textbf{\bibinfo{volume}{6}},
  \bibinfo{pages}{242} (\bibinfo{year}{2011}).

\bibitem[{\citenamefont{Wiseman}(1994)}]{WisemanPRA1994}
\bibinfo{author}{\bibfnamefont{H.~M.} \bibnamefont{Wiseman}},
  \bibinfo{journal}{Phys. Rev. A} \textbf{\bibinfo{volume}{49}},
  \bibinfo{pages}{2133} (\bibinfo{year}{1994}).

\bibitem[{\citenamefont{Piilo et~al.}(2008)\citenamefont{Piilo, Maniscalco,
  H\"ark\"onen, and Suominen}}]{PiiloPRL2008}
\bibinfo{author}{\bibfnamefont{J.}~\bibnamefont{Piilo}},
  \bibinfo{author}{\bibfnamefont{S.}~\bibnamefont{Maniscalco}},
  \bibinfo{author}{\bibfnamefont{K.}~\bibnamefont{H\"ark\"onen}},
  \bibnamefont{and} \bibinfo{author}{\bibfnamefont{K.-A.}
  \bibnamefont{Suominen}}, \bibinfo{journal}{Phys. Rev. Lett.}
  \textbf{\bibinfo{volume}{100}}, \bibinfo{pages}{180402}
  (\bibinfo{year}{2008}).

\bibitem[{\citenamefont{Kofman and Kurizki}(2000)}]{KofmanNature2000}
\bibinfo{author}{\bibfnamefont{A.~G.} \bibnamefont{Kofman}} \bibnamefont{and}
  \bibinfo{author}{\bibfnamefont{G.}~\bibnamefont{Kurizki}},
  \bibinfo{journal}{Nature} \textbf{\bibinfo{volume}{405}},
  \bibinfo{pages}{546} (\bibinfo{year}{2000}).

\end{thebibliography}

\end{document}


\title{Supplementary materials for \textquotedblleft Engineering nuclear spin
dynamics with optically pumped nitrogen-vacancy center\textquotedblright}
\author{Ping Wang }
\affiliation{Hefei National Laboratory for Physics Sciences at Microscale and Department of
Modern Physics, University of Science and Technology of China, Hefei, Anhui
230026, China}
\affiliation{Beijing Computational Science Research Center, Beijing 100084, China}
\author{Jiangfeng Du}
\affiliation{Hefei National Laboratory for Physics Sciences at Microscale and Department of
Modern Physics, University of Science and Technology of China, Hefei, Anhui
230026, China}
\author{Wen Yang}
\affiliation{Beijing Computational Science Research Center, Beijing 100084, China}
\email{wenyang@csrc.ac.cn}
\maketitle

Section I provides a detailed derivation of Eqs. (2)-(6) of the main text.
Sec. II provides a perturbative, explicit expression for the nuclear spin
transition rate $W_{\mathbf{p}\leftarrow\mathbf{m}}$, which can be easily
applied to a given experimental setup. Sec. III details the NV Hamiltonian,
NV-induced nuclear spin transition rates, and calculation of NV fluorescence
under the CPT condition \cite{ToganNature2011}.

\section{Derivation of nuclear spin equations of motion}

The starting point is
\begin{align}
\dot{\rho}_{\mathbf{m},\mathbf{n}}(t)  &  =\mathcal{L}_{\mathbf{m},\mathbf{n}%
}\hat{\rho}_{\mathbf{m},\mathbf{n}}(t)-i\frac{\{\hat{\rho}_{\mathbf{m}%
,\mathbf{n}}(t),\delta\hat{K}_{\mathbf{m},\mathbf{n}}\}}{2}-i\langle
\mathbf{m}|[\hat{V}(t),\hat{\rho}(t)]|\mathbf{n}\rangle,\label{EOM_RHO}\\
\dot{p}_{\mathbf{m},\mathbf{n}}(t)  &  =-i\operatorname*{Tr}\nolimits_{e}%
\frac{\{\hat{\rho}_{\mathbf{m},\mathbf{n}}(t),\delta\hat{K}_{\mathbf{m}%
,\mathbf{n}}\}}{2}-i\operatorname*{Tr}\nolimits_{e}\langle\mathbf{m}|[\hat
{V}(t),\hat{\rho}(t)]|\mathbf{n}\rangle. \label{EOM_P}%
\end{align}
On the coarse grained time scale $T_{e}\ll\Delta t\ll T_{1},T_{2}$, we treat
$\{p_{\mathbf{m},\mathbf{n}}(t)\}$ as slow variables, others as fast
variables, and $\delta\hat{K}_{\mathbf{m},\mathbf{n}}$ and $\hat{V}(t)$ as
first-order small quantities. Correspondingly, we decompose $\hat{\rho
}(t)=\hat{\rho}^{(0)}(t)+\hat{\rho}^{(1)}(t)+\cdots$, where $\hat{\rho}%
^{(k)}(t)$ is a $k$th-order small quantity. Since we always work in the
nuclear spin interaction picture, the zeroth-order dynamics vanishes:
$(\dot{p})_{0}=0$.

The first-order dynamics $(\dot{p}_{\mathbf{m},\mathbf{n}})_{1}$ [Eq. (2) of
the main text] is obtained from Eq. (\ref{EOM_P}) by replacing $\hat{\rho}(t)$
with $\hat{\rho}^{(0)}(t)=\sum_{\mathbf{m},\mathbf{n}}|\mathbf{m}%
\rangle\langle\mathbf{n}|\hat{\rho}_{\mathbf{m},\mathbf{n}}^{(0)}(t)$,
obtained by coarse-graining%
\[
\dot{\rho}_{\mathbf{m},\mathbf{n}}^{(0)}(t)=\mathcal{L}_{\mathbf{m}%
,\mathbf{n}}\hat{\rho}_{\mathbf{m},\mathbf{n}}^{(0)}(t)
\]
for an interval $T_{e}\ll\Delta t\ll T_{1},T_{2}$. Here $\hat{P}%
_{\mathbf{m},\mathbf{n}}$ is the normalized electron steady state determined
by $\mathcal{L}_{\mathbf{m,n}}\hat{P}_{\mathbf{m,n}}=0$. The second-order
dynamics $(\dot{p}_{\mathbf{m},\mathbf{n}})_{2}$ is obtained from Eq.
(\ref{EOM_P}) by replacing $\hat{\rho}(t)$ with $\hat{\rho}^{(1)}(t)$, the
solution to%
\[
(\dot{\rho}_{\mathbf{m},\mathbf{n}}^{(0)})_{1}+\dot{\rho}_{\mathbf{m}%
,\mathbf{n}}^{(1)}=\mathcal{L}_{\mathbf{m},\mathbf{n}}\hat{\rho}%
_{\mathbf{m},\mathbf{n}}^{(1)}-i\frac{\{\hat{\rho}_{\mathbf{m},\mathbf{n}%
}^{(0)},\delta\hat{K}_{\mathbf{m},\mathbf{n}}\}}{2}-i\langle\mathbf{m}%
|[\hat{V},\hat{\rho}^{(0)}]\mathbf{n}\rangle,
\]
where $(\dot{\rho}_{\mathbf{m},\mathbf{n}}^{(0)})_{1}=\hat{P}_{\mathbf{m}%
,\mathbf{n}}(\dot{p}_{\mathbf{m},\mathbf{n}})_{1}$ comes from the first-order
evolution $(\dot{p}_{\mathbf{m},\mathbf{n}})_{1}$. Substituting into the above
equation and coarse-graining for an interval $1/\gamma_{e}\ll\Delta t\ll
T_{1},T_{2}$ gives%
\begin{align*}
\hat{\rho}_{\mathbf{m},\mathbf{n}}^{(1)}=ip_{\mathbf{m},\mathbf{n}}%
\mathcal{L}_{\mathbf{m},\mathbf{n}}^{-1}\frac{\{\hat{P}_{\mathbf{m}%
,\mathbf{n}},\delta\tilde{K}_{\mathbf{m},\mathbf{n}}\}}{2}  &  +i\sum
_{\mathbf{p}}(\mathcal{L}_{\mathbf{m},\mathbf{n}}+i\omega_{\mathbf{m}%
,\mathbf{p}})^{-1}(\hat{V}_{\mathbf{m},\mathbf{p}}\hat{P}_{\mathbf{p}%
,\mathbf{n}}-\langle\hat{V}_{\mathbf{m},\mathbf{p}}\rangle_{\mathbf{p}%
,\mathbf{n}}\hat{P}_{\mathbf{m},\mathbf{n}})p_{\mathbf{p},\mathbf{n}}\\
&  -i\sum_{\mathbf{p}}(\mathcal{L}_{\mathbf{m},\mathbf{n}}+i\omega
_{\mathbf{p},\mathbf{n}})^{-1}(\hat{P}_{\mathbf{m,p}}\hat{V}_{\mathbf{p}%
,\mathbf{n}}-\langle\hat{V}_{\mathbf{p},\mathbf{n}}\rangle_{\mathbf{m}%
,\mathbf{p}}\hat{P}_{\mathbf{m},\mathbf{n}})p_{\mathbf{m,p}},
\end{align*}
where $(\mathcal{L}_{\mathbf{m},\mathbf{n}}+i\omega_{\mathbf{m},\mathbf{p}%
})^{-1}\equiv-\int_{0}^{\infty}e^{(\mathcal{L}_{\mathbf{m},\mathbf{n}}%
+i\omega_{\mathbf{m},\mathbf{p}})t}dt$, $\mathcal{L}_{\mathbf{m},\mathbf{n}%
}^{-1}\equiv-\int_{0}^{\infty}e^{\mathcal{L}_{\mathbf{m},\mathbf{n}}t}%
dt=\lim_{\omega\rightarrow0}(\mathcal{L}_{\mathbf{m},\mathbf{n}}+i\omega
)^{-1}$, $\left\langle \cdots\right\rangle _{\mathbf{m},\mathbf{n}}%
\equiv\operatorname*{Tr}_{e}(\cdots)\hat{P}_{\mathbf{m},\mathbf{n}}$, and we
have assumed $\hat{V}_{\mathbf{m},\mathbf{n}}(t)=\hat{V}_{\mathbf{m}%
,\mathbf{n}}e^{-i\omega_{\mathbf{m},\mathbf{n}}t}$. Then replacing $\hat{\rho
}(t)$ with $\hat{\rho}^{(1)}(t)$ in Eq. (\ref{EOM_P}) gives the desired
second-order nuclear spin dynamics. For $\mathbf{m}=\mathbf{n}$, neglecting
the coupling of $p_{\mathbf{m},\mathbf{m}}$ to other $p_{\mathbf{p}%
,\mathbf{q}}$ $(\mathbf{p}\neq\mathbf{q})$, which amounts to neglecting the
small second-order corrections to $\langle\hat{V}(t)\rangle$ and
electron-mediated nuclear spin interactions (they induce nuclear spin
diffusion and depolarization, which can be included phenomenologically at the
end of the derivation), we obtain Eq. (3) of the main text. The transition
rate%
\[
W_{\mathbf{p}\leftarrow\mathbf{m}}\equiv-2\operatorname{Re}\operatorname*{Tr}%
\nolimits_{e}\hat{V}_{\mathbf{m},\mathbf{p}}(\mathcal{L}_{\mathbf{p,m}%
}+i\omega_{\mathbf{p,m}})^{-1}(\hat{V}_{\mathbf{p},\mathbf{m}}\hat
{P}_{\mathbf{m},\mathbf{m}}-\langle\hat{V}_{\mathbf{p},\mathbf{m}}%
\rangle_{\mathbf{m},\mathbf{m}}\hat{P}_{\mathbf{p,m}}).
\]
Using $(\mathcal{L}_{\mathbf{p,m}}+i\omega_{\mathbf{p,m}})^{-1}\hat
{P}_{\mathbf{p,m}}=\hat{P}_{\mathbf{p,m}}/(i\omega_{\mathbf{p,m}})$ and
$P_{\mathbf{p},\mathbf{m}}\approx P_{\mathbf{m},\mathbf{m}}$ ($|\mathbf{p}%
\rangle$ and $|\mathbf{m}\rangle$ differs only by a single nuclear spin flip),
we see that the term $\propto\hat{P}_{\mathbf{p,m}}$ is $(2/\omega
_{\mathbf{p,m}})\operatorname{Im}\langle\hat{V}_{\mathbf{m},\mathbf{p}}%
\rangle_{\mathbf{p},\mathbf{m}}\langle\hat{V}_{\mathbf{p},\mathbf{m}}%
\rangle_{\mathbf{m},\mathbf{m}}\approx0$ and hence recover Eq. (4) of the main
text. For $\mathbf{m}\neq\mathbf{n}$, neglecting the coupling of
$p_{\mathbf{m},\mathbf{n}}$ to other variables and second-order self-energy
corrections, which amounts to neglecting the small second-order corrections to
$\langle\hat{V}(t)\rangle$ and electron-mediated nuclear spin interactions, we
obtain Eqs. (5) and (6) of the main text, where%

\[
W_{\mathbf{p}\leftarrow\mathbf{m|n}}\equiv-2\operatorname{Re}%
\operatorname*{Tr}\nolimits_{e}\hat{V}_{\mathbf{m,p}}(\mathcal{L}%
_{\mathbf{p,n}}+i\omega_{\mathbf{p,m}})^{-1}(\hat{V}_{\mathbf{p,m}%
}P_{\mathbf{m,n}}-\langle\hat{V}_{\mathbf{p,m}}\rangle_{\mathbf{m},\mathbf{n}%
}P_{\mathbf{p,n}}).
\]

\section{Perturbative, explicit expression for $W_{\mathbf{p}\leftarrow
\mathbf{m}}$}

If $\langle\mathbf{p}|\hat{V}(t)|\mathbf{m}\rangle$ has multiple frequencies
that are widely separated compared with the nuclei transition/dephasing rates,
then $W_{\mathbf{p}\leftarrow\mathbf{m}}$ is the sum of contributions from
different frequency components. So we consider the case $\langle
\mathbf{p}|\hat{V}(t)|\mathbf{m}\rangle=\hat{F}e^{-i\omega t}$ with a single
frequency. Equation (4)\ of the main text gives\quad$W_{\mathbf{p}%
\leftarrow\mathbf{m}}=-2\mathrm{Re}\mathrm{Tr}_{e}\hat{F}^{\dagger}%
\mathcal{G}\hat{F}\hat{P}$,\quad where $\mathcal{G}\equiv(i\omega
+\mathcal{L})^{-1}$, $\hat{P}\equiv\hat{P}_{\mathbf{m},\mathbf{m}}$, and
$\mathcal{L}\equiv\mathcal{L}_{\mathbf{p},\mathbf{m}}$. The key is to
calculate $\mathcal{G}$ perturbatively by dividing $\mathcal{L}(\bullet
)\equiv-i[\hat{H},\bullet]+\mathcal{L}_{e}(\bullet)$ into the unperturbed
part
\[
\mathcal{L}^{\mathrm{d}}(\bullet)\equiv-i[\hat{H}^{\mathrm{d}},\bullet
]-\frac{1}{2}\sum_{i}\Gamma_{i}\{|i\rangle\langle i|,\bullet\}+\sum_{i}%
\gamma_{ii}|i\rangle\langle i|\bullet|i\rangle\langle i|
\]
and the perturbation
\[
\mathcal{L}^{\mathrm{nd}}(\bullet)\equiv-i[\hat{H}^{\mathrm{nd}},\bullet
]+\sum_{f\neq i}\gamma_{fi}|f\rangle\langle f|\langle i|\bullet|i\rangle,
\]
where $\hat{H}^{\mathrm{d}}=\sum_{i}\varepsilon_{i}|i\rangle\langle i|$
($\hat{H}^{\mathrm{nd}}$) is the diagonal (off-diagonal) part of $\hat{H}$ and
$\Gamma_{i}\equiv\sum_{f}\gamma_{fi}$. When $||\mathcal{L}^{\mathrm{d}%
}+i\omega||\gg||\mathcal{L}^{\mathrm{nd}}||$, an explicit expression for
$W_{\mathbf{p}\leftarrow\mathbf{m}}$ is obtained by replacing $\mathcal{G}$ by
$\mathcal{G}^{\mathrm{d}}-\mathcal{G}^{\mathrm{d}}\mathcal{L}^{\mathrm{nd}%
}\mathcal{G}^{\mathrm{d}}$ and using $\mathcal{G}^{\mathrm{d}}|k\rangle\langle
j|\equiv(\mathcal{L}^{\mathrm{d}}+i\omega)^{-1}|k\rangle\langle j|=|k\rangle
\langle j|(i/z_{k,j})$ with $z_{k,j}\equiv\varepsilon_{k}-\varepsilon
_{j}-\omega-i(\Gamma_{k}+\Gamma_{j}-2\gamma_{k,k}\delta_{k,j})/2$. For
$\hat{F}=\sum_{f\neq i}V_{fi}|f\rangle\langle i|$, we obtain $W_{\mathbf{p}%
\leftarrow\mathbf{m}}$ as the sum of the golden rule part
\begin{equation}
W_{\mathbf{p}\leftarrow\mathbf{m}}^{\mathrm{golden}}=2\sum_{i,i^{\prime}%
,f}\operatorname{Im}\frac{V_{f,i^{\prime}}^{\ast}V_{f,i}P_{i,i^{\prime}}%
}{z_{f,i^{\prime}}} \label{WPM_GOLDEN}%
\end{equation}
and the coherent part%

\begin{equation}
W_{\mathbf{p}\leftarrow\mathbf{m}}^{\mathrm{coh}}=2\operatorname{Im}%
\sum_{i,i^{\prime},f,f^{\prime}}\left(  \frac{H_{i^{\prime},f^{\prime}%
}^{\mathrm{nd}}V_{f,f^{\prime}}^{\ast}}{z_{f,f^{\prime}}}-\frac{V_{f^{\prime
},i^{\prime}}^{\ast}H_{f^{\prime},f}^{\mathrm{nd}}}{z_{f^{\prime},i^{\prime}}%
}\right)  \frac{V_{f,i}P_{i,i^{\prime}}}{z_{f,i^{\prime}}}, \label{WPM_COH}%
\end{equation}
where $O_{i,i^{\prime}}\equiv\langle i|\hat{O}|i^{\prime}\rangle$.

\section{NV Hamiltonian, nuclear spin transition rates, and NV fluorescence
under CPT}

\subsection{NV Hamiltonian under CPT}

The NV Hamiltonian for the coherent population trapping (CPT) experiment has
been discussed in \cite{ToganNature2011}. Here we reproduce it with greater
detail using our own notations. The strain effect is neglected as we have
confirmed that it produces a small influence on the nuclear spin polarization
and noise suppression. The NV states of relevance include 3 ground triplet
states $|0\rangle$ (energy $0$), $|\pm1\rangle$ (energy $D_{\mathrm{gs}}$), 6
excited triplet states $|E_{y}\rangle,|E_{x}\rangle$ (energy $\varepsilon
_{E_{x}}=\varepsilon_{E_{y}}$)$,|E_{1}\rangle,|E_{2}\rangle$ (energy
$\varepsilon_{E_{1}}=\varepsilon_{E_{2}}$), $|A_{1}\rangle$ (energy
$\varepsilon_{A_{1}}$)$,|A_{2}\rangle$ (energy $\varepsilon_{A_{2}}$), and one
metastable singlet $|S\rangle$ (energy $\varepsilon_{S}$). A linearly
polarized laser with electric field $\mathbf{E}_{1}e^{-i\omega_{1}t}/2+c.c.$
and frequency $\omega_{1}=\varepsilon_{A_{1}}-D_{\mathrm{gs}}$ resonantly
excites the ground states $|\pm1\rangle$ to the excited state $|A_{1}\rangle$
and, at the same time, off-resonantly excited $|\pm1\rangle$ to the excited
state $|A_{2}\rangle$ with detuning $\Delta=\varepsilon_{A_{2}}-\varepsilon
_{A_{1}}$. Another linearly polarized laser with electric field $\mathbf{E}%
_{2}e^{-i\omega_{2}t}/2+c.c.$ and frequency $\omega_{2}=\varepsilon_{E_{x}}$
resonantly excites $|0\rangle$ to $|E_{y}\rangle$. The relevant optical
transition matrix element of the electric dipole operator $\mathbf{d}%
\equiv-e\mathbf{r}$ are $\langle A_{1}|\mathbf{d}|\pm1\rangle=\pm
d_{a,E}\mathbf{e}_{\pm}/(2\sqrt{2})$, $\langle A_{2}|\mathbf{d}|\pm
1\rangle=id_{a,E}\mathbf{e}_{\pm}/(2\sqrt{2})$, and $\langle E_{y}%
|\mathbf{d}|0\rangle=d_{a,E}\mathbf{e}_{x}/\sqrt{2}$, where $\mathbf{e}_{\pm
}\equiv\mathbf{e}_{x}\pm i\mathbf{e}_{y}$ and $d_{a,E}$ is the reduced matrix
element of the electric dipole moment \cite{DohertyNJP2011}. Thus the laser
coupling Hamiltonian
\[
\hat{H}_{c}(t)=\frac{\Omega_{A}}{2}(e^{i\phi}\hat{\sigma}_{A_{1},+1}%
-e^{-i\phi}\hat{\sigma}_{A_{1},-1}+ie^{i\phi}\hat{\sigma}_{A_{2}%
,+1}+ie^{-i\phi}\hat{\sigma}_{A_{2},-1})e^{-i\omega_{1}t}+\frac{\Omega_{E}}%
{2}e^{-i\omega_{2}t}\hat{\sigma}_{E_{y},0}+h.c.,
\]
where we have defined $\hat{\sigma}_{ij}\equiv|i\rangle\langle j|$,
$\Omega_{E}=\mathbf{E}_{2}\cdot\langle E_{y}|\mathbf{d}|0\rangle
=E_{2,x}d_{a,E}/\sqrt{2}$, and $\Omega_{A}e^{i\phi}\equiv\mathbf{E}_{1}%
\cdot\langle A_{1}|\mathbf{d}|+1\rangle=d_{aE}(E_{1,x}+iE_{1,y})/(2\sqrt{2})$.
Defining the bright state $|b\rangle=(e^{-i\phi}|+1\rangle-e^{i\phi}%
|-1\rangle)/\sqrt{2}$ and dark state $|d\rangle=(e^{-i\phi}|+1\rangle
+e^{i\phi}|-1\rangle)/\sqrt{2}$, the laser coupling Hamiltonian simplifies to
\[
\hat{H}_{c}(t)=\frac{\Omega_{A}}{\sqrt{2}}(\hat{\sigma}_{A_{1},b}+i\hat
{\sigma}_{A_{2},d})e^{-i\omega_{1}t}+\frac{\Omega_{E}}{2}e^{-i\omega_{2}t}%
\hat{\sigma}_{E_{y},0}+h.c..
\]
In the rotating frame $|\psi(t)\rangle$ connected to the laboratory frame
$|\psi_{\mathrm{lab}}(t)\rangle$ via $|\psi(t)\rangle=e^{i\hat{H}_{0}t}%
|\psi_{\mathrm{lab}}(t)\rangle$ with $\hat{H}_{0}=\varepsilon_{E_{x}}%
(\hat{\sigma}_{E_{x},E_{x}}+\hat{\sigma}_{E_{y},E_{y}})+\varepsilon_{E_{1}%
}(\hat{\sigma}_{E_{1},E_{1}}+\hat{\sigma}_{E_{2},E_{2}})+\varepsilon_{A_{1}%
}(\hat{\sigma}_{A_{1},A_{1}}+\hat{\sigma}_{A_{2},A_{2}})+\varepsilon_{S}%
\hat{\sigma}_{S,S}+D_{\mathrm{gs}}(\hat{\sigma}_{1,1}+\hat{\sigma}_{-1,-1})$,
the NV Hamiltonian is%
\[
\hat{H}_{e}=\omega_{e}\hat{S}_{g}^{z}+\Delta\hat{\sigma}_{A_{2},A_{2}}%
+\frac{\Omega_{A}}{\sqrt{2}}(\hat{\sigma}_{A_{1},b}+i\hat{\sigma}_{A_{2}%
,d}+h.c.)+\frac{\Omega_{E}}{2}(\hat{\sigma}_{E_{y},0}+\hat{\sigma}_{0,E_{y}%
}),
\]
where we have included the ground state Zeeman term $g_{e}\mu_{B}B\hat{S}%
_{g}^{z}\equiv\omega_{e}\hat{S}_{g}^{z}$ of an external magnetic field $B$
along the $z$ (N-V) axis. We also include the dissipation of the NV center in
the Lindblad form, including the spontaneous emission from $|A_{1}\rangle$ and
$|A_{2}\rangle$ to $|\pm1\rangle$ (rate $\gamma/2)$, from $|E_{y}\rangle$ to
$|0\rangle$ (rate $\gamma$), and non-radiative intersystem crossing from
$|A_{1}\rangle$ to $|S\rangle$ (rate $\gamma_{s1}$), from $|A_{2}\rangle$ to
$|S\rangle$ (rate $\gamma_{s2}$), from $|S\rangle$ to $|0\rangle$ (rate
$\gamma_{s1}$), from $|E_{y}\rangle$ to $|\pm1\rangle$ (rate $\gamma_{ce}$),
and pure dephasing $\gamma_{\varphi}$ for each excited state.

\subsection{NV-induced nuclear spin transition rate}

To calculate the NV-induced nuclear spin transition rates, we need to
determine the NV steady state $\hat{P}(h)$ from $-i[\hat{H}_{e}+\hat{S}%
_{g}^{z}h,\hat{P}]+\mathcal{L}_{e}\hat{P}(h)=0$, where $\hat{H}_{e}+\hat
{S}_{g}^{z}h=\hat{H}_{e}|_{\omega_{e}\rightarrow\delta\equiv\omega_{e}+h}$ and
$\delta\equiv\omega_{e}+h$ is the two-photon detuning. The calculation is
straightforward and it turns out that relevant matrix elements of $\hat{P}(h)$
can be expressed in terms of the $|E_{y}\rangle$ population $\langle
E_{y}|\hat{P}(h)|E_{y}\rangle$, e.g., $\langle S|\hat{P}(h)|S\rangle
=(2\gamma_{ce}/\gamma_{s})\langle E_{y}|\hat{P}(h)|E_{y}\rangle$,
$\langle0|\hat{P}(h)|0\rangle=\left[  1+(\gamma+2\gamma_{ce})/W_{E}\right]
\langle E_{y}|\hat{P}(h)|E_{y}\rangle$, and $\langle0|\hat{P}(h)|E_{y}%
\rangle=[i(\gamma+2\gamma_{ce})/\Omega_{E}]\langle E_{y}|\hat{P}%
(h)|E_{y}\rangle$, where $W_{E}\equiv\Omega_{E}^{2}/(2\gamma_{ce}%
+\gamma+\gamma_{\varphi})$ is the transition rate between $|0\rangle$ and
$|E_{y}\rangle$, $W_{A}=\Omega_{A}^{2}/(\gamma+\gamma_{s1}+\gamma_{\varphi})$
is the transition rate between $|\pm1\rangle$ and $|A_{1}\rangle$, and
$\eta_{1}\equiv\gamma_{ce}/\gamma_{s1}$. We treat the off-resonant
$|\pm1\rangle\rightarrow|A_{2}\rangle$ excitation perturbatively. Up to zeroth
order, near the two-photon resonance $\delta\approx0$, we neglect
$O(\delta^{4})$ and high order terms and obtain%
\[
\langle E_{y}|\hat{P}^{(0)}(h)|E_{y}\rangle\approx D_{0}\frac{\delta^{2}%
}{\delta^{2}+\delta_{0}^{2}},
\]
where%
\begin{align*}
D_{0} &  \approx\frac{1}{\frac{2}{\eta_{2}}+2\eta_{1}\frac{\gamma+\gamma_{s1}%
}{W_{A}}+\frac{\gamma+2\gamma_{ce}}{W_{E}}},\\
\delta_{0}^{2} &  \approx\frac{\eta_{1}\eta_{2}}{4}\frac{W_{A}^{2}}{\eta
_{1}\eta_{2}+\frac{W_{A}}{\gamma+\gamma_{s1}}\left(  1+\frac{\eta_{2}}{2}%
\frac{\gamma+2\gamma_{ce}}{W_{E}}\right)  },
\end{align*}
with $\eta_{2}\equiv\gamma_{s}/(\gamma_{s}+\gamma_{ce})$ and we have used
$\eta_{1}\ll1$. We find that only $|E_{y}\rangle,|0\rangle$, and $|d\rangle$
are significantly populated since $\gamma_{ce}\ll\gamma_{s},\gamma_{s1}%
,\gamma$. Under saturated pumping $W_{E}\gg\gamma$, the leading order
correction from the off-resonant excitation to $|A_{2}\rangle$ is%
\[
\langle E_{y}|\hat{P}^{(2)}(h)|E_{y}\rangle\approx\frac{1}{2\eta_{1}}%
\frac{W_{A_{2}}}{\gamma+\gamma_{s1}}(1-2\langle E_{y}|\hat{P}^{(0)}%
(h)|E_{y}\rangle),
\]
where $W_{A_{2}}=(\Omega_{A}^{2}/2)(\gamma+\gamma_{s2}+\gamma_{\varphi
})/\Delta^{2}$ is the off-resonant transition rate between $|\pm1\rangle$ and
$|A_{2}\rangle$ and we have used $\gamma_{s2}\ll\gamma,\gamma_{s1}$. Now the
$|E_{y}\rangle$ population $\langle E_{y}|\hat{P}^{(0)}(h)|E_{y}%
\rangle+\langle E_{y}|\hat{P}^{(2)}(h)|E_{y}\rangle$ no longer vanish at
$\delta=0$. This limits the efficiency of $^{14}$N nuclear spin cooling (in
agreement with the expectation of Ref. \cite{ToganNature2011}) and $^{13}$C
nuclear spin noise suppression.

Next we use Eqs. (\ref{WPM_GOLDEN}) and (\ref{WPM_COH}) [with $\hat
{P}_{\mathbf{m},\mathbf{m}}$ replaced by $\hat{P}(h)$ and neglecting the
dependence of $\mathcal{L}_{\mathbf{p,m}}$ on $\mathbf{p}$ and $\mathbf{m}$]
to calculate the $^{14}$N and $^{13}$C nuclear spin transition rate. The
transition rate $W_{m_{0}+1\leftarrow m_{0}}(h)=W_{m_{0}-1\leftarrow m_{0}%
}(h)$ for $^{14}$N from $|m_{0}\rangle$ to $|m_{0}\pm1\rangle\propto\hat
{I}_{0}^{\pm}|m_{0}\rangle$ is dominated by the following contributions from
different NV transitions: $A_{g}^{2}\chi_{g}\langle E_{y}|\hat{P}%
(h)|E_{y}\rangle$ from $|0\rangle\rightarrow|\pm1\rangle$ and $A_{e}^{2}%
(\sum_{f}\chi_{f})\langle E_{y}|\hat{P}(h)|E_{y}\rangle$ from $|E_{y}%
\rangle\rightarrow|f\rangle$, where, $f$ runs over $A_{1},A_{2},E_{1},E_{2}$
states, $\chi_{g}\equiv(\gamma+2\gamma_{ce})/D_{\mathrm{gs}}^{2}$, and
$\chi_{f}\equiv(1/4)(\Gamma_{f}+\gamma_{\varphi})/(\varepsilon_{E_{y}%
}-\varepsilon_{f})^{2}$, with $\varepsilon_{f}$ the energy of $|f\rangle$ in
the laboratory frame. These transition rates differ from the phenomenlogical
expression $A_{e}\langle E_{y}|\hat{P}|E_{y}\rangle$ in Ref.
\cite{ToganNature2011}. Similarly, the transition rate $W_{m_{n}\pm1\leftarrow
m_{n}}$ of the $n$th $^{13}$C nucleus from $|m_{n}\rangle$ to $|m_{n}%
\pm1\rangle\propto\hat{I}_{n}^{\pm}|m_{n}\rangle$ is dominated by the
following contributions: $\chi_{g}(|A_{n,-,-}|^{2}+|A_{n,+,-}|^{2})\langle
E_{y}|\hat{P}(h)|E_{y}\rangle/8$ from $|0\rangle\rightarrow|\pm1\rangle$,
$|A_{n,Y,-}|^{2}(\chi_{A_{1}}+\chi_{E_{1}})\langle E_{y}|\hat{P}%
(h)|E_{y}\rangle/2$ from $|E_{y}\rangle\rightarrow|A_{1}\rangle,|E_{1}\rangle
$, and $|A_{n,X,-}|^{2}(\chi_{A_{2}}+\chi_{E_{2}})\langle E_{y}|\hat
{P}(h)|E_{y}\rangle/2$ from $|E_{y}\rangle\rightarrow|A_{2}\rangle
,|E_{2}\rangle$, where $A_{n,\alpha,\beta}\equiv\mathbf{e}_{n,\alpha}%
\cdot\mathbf{A}_{n}\cdot\mathbf{e}_{n,\beta}$ and $\mathbf{e}_{n,\pm}%
\equiv\mathbf{e}_{n,X}\pm i\mathbf{e}_{n,Y}$.

\subsection{Calculation of NV fluorescence}

To compare with the experimentally observed suppressed $^{13}$C nuclear spin
noise, we first set $\omega_{e}=0.18\mathrm{MHz}$ to obtain the steady nuclear
state populations $\{p_{\mathbf{m},\mathbf{m}}\}$ and then calculate the
nuclear-state-dependent NV population $\langle E_{y}|\hat{P}_{\mathbf{m}%
,\mathbf{m}}|E_{y}\rangle$ at the readout magnetic field $\omega_{\mathrm{re}%
}=g_{e}\mu_{B}B_{\mathrm{re}}$. The fluorescence before post-selection is
proportional to $\sum_{\mathbf{m}}p_{\mathbf{m},\mathbf{m}}\langle E_{y}%
|\hat{P}_{\mathbf{m},\mathbf{m}}|E_{y}\rangle$, while the fluorescence after
post-selection is proportional to $\sum_{\mathbf{m}}p_{\mathbf{m},\mathbf{m}%
}\langle E_{y}|\hat{P}_{\mathbf{m},\mathbf{m}}|E_{y}\rangle e^{-\epsilon\gamma
T_{\mathrm{cond}}\langle E_{y}|\hat{P}_{\mathbf{m},\mathbf{m}}|E_{y}\rangle}$,
where $\epsilon$ is the photon collection efficiency and $T_{\mathrm{cond}}$
is the photon collection time.

\bibliographystyle{apsrev}
\bibliography{myLiterature}